\documentclass[10pt,journal,compsoc]{IEEEtran}

\def\reviewpass{5.0.0}

\newif \ifDraft \Drafttrue
\Draftfalse

\newif \ifOpen \Opentrue
\Opentrue

\ifCLASSOPTIONcompsoc
  \usepackage[nocompress]{cite}
\else
  \usepackage{cite}
\fi

\ifCLASSINFOpdf
\else
\fi

\usepackage{array}

\newcommand{\Comment}[1]{\textbf{\textsl{#1}}}

\newcommand{\todo}[1]{}
\newcommand{\gernot}[1]{}
\newcommand{\nils}[1]{}
\newcommand{\luca}[1]{}
\newcommand{\moritz}[1]{}

\newcommand{\riscv}{\mbox{RISC-V}}

\newcommand{\fencet}{\mbox{\texttt{fence.t}}}

\newcommand{\cspad}{\mbox{\texttt{cspad}}}

\newcommand{\firstflush}{\textsc{Flush$_1$}}

\newcommand{\fullflush}{\textsc{Flush$_2$}}

\usepackage[acronym]{glossaries}

\usepackage[dvipsnames]{xcolor}

\usepackage{tikz}
\usetikzlibrary{positioning}
\usetikzlibrary{calc}
\usetikzlibrary{patterns}

\usepackage{pgfplots}
\pgfplotsset{compat=newest}

\usepackage{hyperref}

\usepackage[english]{babel}

\usepackage[style= base]{subcaption}

\usepackage[binary-units=true,detect-all]{siunitx}
\DeclareSIUnit\GE{GE}

\usepackage{booktabs}
\usepackage{etoolbox}
\robustify\bfseries

\usepackage{multirow}

\usepackage{balance}

\ifDraft
  \newwatermark[
    allpages,
    color = red!50,
    angle = 90,
    scale = 1,
    xpos = -0.5\paperwidth+5mm,
    ypos = 0.5\paperheight-4cm
  ]{v\reviewpass}
\else
  \renewcommand{\Comment}[1]{\relax}
\fi

\usepackage{amsthm}
\theoremstyle{definition}
\newtheorem{requirement}{Requirement}

\newtheorem{step}{Step}

\usepackage[normalem]{ulem}

\ifDraft

\else

\fi

\newacronym{asic}{ASIC}{Application-Specific Integrated Circuit}
\newacronym{cpu}{CPU}{central processing unit}
\newacronym{isa}{ISA}{instruction set architecture}
\newacronym[shortplural={OSes}]{os}{OS}{operating system}
\newacronym{fp}{FP}{floating-point}
\newacronym{fpu}{FPU}{floating-point unit}
\newacronym{axi}{AXI}{Advanced eXtensible Interface}
\newacronym[firstplural=systems on chip (SoCs)]{soc}{SoC}{system on chip}
\newacronym{hwpe}{HWPE}{hardware processing element}
\newacronym[firstplural=multiprocessor systems on chip (MPSoCs)]{mpsoc}{MPSoC}{multiprocessor system on a chip}
\newacronym{mpam}{MPAM}{memory partitioning and monitoring}
\newacronym{mba}{MBA}{memory bandwidth allocation}
\newacronym{qos}{QoS}{quality of service}
\newacronym{wcet}{WCET}{worst-case execution time}
\newacronym[firstplural=networks on chip (NoCs)]{noc}{NoC}{network on chip}
\newacronym{rpc}{RPC}{reduced pin count}
\newacronym{tlb}{TLB}{translation lookaside buffer}
\newacronym{bht}{BHT}{branch history table}
\newacronym{btb}{BTB}{branch target buffer}
\newacronym{lru}{LRU}{least-recently-used}
\newacronym{plru}{pseudo-LRU}{pseudo-least-recently-used}
\newacronym{lfsr}{LFSR}{linear-feedback shift register}
\newacronym{clint}{CLINT}{core-local interrupt controller}
\newacronym{csr}{CSR}{control and status register}
\newacronym{fsm}{FSM}{finite-state machine}

\begin{document}
\title{Systematic Prevention of On-Core Timing Channels by Full Temporal Partitioning}

\author{Nils~Wistoff,~\IEEEmembership{Student Member,~IEEE,}
        Moritz~Schneider,~\IEEEmembership{Student Member,~IEEE,}
        Frank~K.~G\"urkaynak,
        Gernot~Heiser,~\IEEEmembership{Fellow,~IEEE,}
        and~Luca~Benini,~\IEEEmembership{Fellow,~IEEE}%
\IEEEcompsocitemizethanks{
\IEEEcompsocthanksitem N. Wistoff, F. K. G\"urkaynak, and L. Benini are with the Integrated Systems Laboratory (IIS), ETH Z\"urich, Switzerland.\protect\\
E-mail: \{nwistoff,kgf,lbenini\}@iis.ee.ethz.ch
\IEEEcompsocthanksitem M. Schneider is with the Institute of Information Security, ETH Z\"urich, Switzerland. %
E-mail: moritz.schneider@inf.ethz.ch
\IEEEcompsocthanksitem G. Heiser is with UNSW Sydney, Australia. %
E-mail: gernot@unsw.edu.au
\IEEEcompsocthanksitem L. Benini also is with the Department of Electrical, Electronic and Information Engineering (DEI), University of Bologna, Bologna, Italy
}%
\ifOpen
\thanks{\copyright~2022~IEEE. Personal use of this material is permitted. Permission from IEEE must be obtained for all other uses, in any current or future media, including reprinting/republishing this material for advertising or promotional purposes, creating new collective works, for resale or redistribution to servers or lists, or reuse of any copyrighted component of this work in other works.}
\else
\thanks{Manuscript received (MONTH) (DAY), (YEAR); revised (MONTH) (DAY), (YEAR).}
\fi
}

\ifOpen
\ifCLASSOPTIONpeerreview
\markboth{IEEE Transactions on Computers}%
{Systematic Prevention of On-Core Timing Channels by Full Temporal Partitioning}
\else
\markboth{IEEE Transactions on Computers}%
{Wistoff \MakeLowercase{\textit{et al.}}: Systematic Prevention of On-Core Timing Channels by Full Temporal Partitioning}
\fi
\else
\ifCLASSOPTIONpeerreview
\markboth{IEEE Transactions on Computers,~Vol.~(Vol), No.~(No), Month~Year}%
{Systematic Prevention of On-Core Timing Channels by Full Temporal Partitioning}
\else
\markboth{IEEE Transactions on Computers,~Vol.~(Vol), No.~(No), Month~Year}%
{Wistoff \MakeLowercase{\textit{et al.}}: Systematic Prevention of On-Core Timing Channels by Full Temporal Partitioning}
\fi
\fi

\IEEEtitleabstractindextext{%
\begin{abstract}
Microarchitectural timing channels enable unwanted information flow across security boundaries, violating fundamental security assumptions.
They leverage timing variations of several state-holding microarchitectural components and have been demonstrated across instruction set architectures and hardware implementations.
Analogously to memory protection, Ge et al.~\cite{Ge2019a} have proposed \emph{time protection} for preventing information leakage via timing channels.
They also showed that time protection calls for hardware support.
This work leverages the open and extensible \riscv{} instruction set architecture (ISA) to introduce the temporal fence instruction \fencet{}, which provides the required mechanisms by clearing vulnerable microarchitectural state and guaranteeing a history-independent context-switch latency.
We propose and discuss three different implementations of \fencet{} and implement them on an experimental version of the seL4 microkernel~\cite{Klein2014_seL4} and CVA6, an open-source, in-order, application class, 64-bit \riscv{} core~\cite{Zaruba2019}.
We find that a complete, systematic, ISA-supported erasure of all non-architectural core components is the most effective implementation while featuring a low implementation effort, a minimal performance overhead of less than \SI{1}{\percent}, and negligible hardware costs.
\end{abstract}

}

\maketitle

\IEEEdisplaynontitleabstractindextext

\IEEEpeerreviewmaketitle

\ifCLASSOPTIONcompsoc
\IEEEraisesectionheading{\section{Introduction}\label{sec:introduction}}
\else
\section{Introduction}
\label{sec:introduction}
\fi

\IEEEPARstart{C}{omputing} systems are trusted with an increasing amount of sensitive information and a large number of safety- and security-critical tasks.
Some examples include personal computing, industrial and military applications, and critical infrastructure.
To prevent unauthorised and malicious access, computer architects have established a set of mechanisms that allow operating systems to isolate concurrent applications through \emph{memory protection}, creating a security boundary.

As the Spectre attacks have prominently demonstrated~\cite{Kocher2018spectre} and several other attacks have confirmed since~\cite{Vanschaik2019ridl,Ren2021uOps}, memory protection is insufficient for a complete isolation of applications.
Concurrently running applications compete for shared hardware resources, which may result in timing variations of one application due to another application's execution.
These timing differences can be leveraged to transfer information between applications, bypassing the security boundary.
We refer to such information channels as \emph{microarchitectural timing channels}.

Ge et al. have proposed supplementing existing memory protection with \emph{time protection} to prevent such timing channels \cite{Ge2019a}.
The key idea is to partition all shared hardware resources---either spatially or temporally---to prevent interference between concurrent applications.
While they show that off-core resources (such as last-level caches) can be efficiently spatially partitioned from software, partitioning on-core  resources requires hardware support beyond that specified in current \glspl{isa}.

In this work, we leverage the open and extensible \riscv{} \gls{isa} to propose a streamlined and ultra-low overhead \gls{isa} extension and hardware implementation for on-core time protection. Specifically, we make the following contributions:

\begin{itemize}
    \item We demonstrate the presence of serious timing channels even in an in-order \riscv{} core and confirm previous claims~\cite{Ge2018} that the current \gls{isa} lacks mechanisms to close them.
    \item We propose the \emph{temporal fence} instruction, \fencet{}, which allows software to partition on-core micrarchitectural resources.
    \item For implementing \fencet{}, we propose \textsc{Microreset}, a systematic erasure of all non-architectural processor state. We compare \textsc{Microreset} to two alternative implementations of \fencet{}: (i) A flush of all well-known vulnerable microarchitectural components, (ii) an exhaustive flush of an extended set of manually identified vulnerable state-holding components.
    \item We propose a context-switch with a history-independent latency, including said reset of microarchitectural state.
    This is of particular (but not exclusive) relevance for a write-back configuration of the L1 data cache.
\end{itemize}

We implement the proposed mechanisms and evaluate their efficacy and costs on CVA6, an open-source, application-class, in-order, 64-bit \riscv{} core~\cite{Zaruba2019} for different cache configurations, and on seL4, an open-source, secure microkernel with formal verification~\cite{Klein2014_seL4}.\footnote{We use an experimental version with time protection support that has not yet been formally verified.}

Using benchmarks, we show that the temporal fence introduces a low performance overhead of less than \SI{1}{\percent}.
The hardware modifications do not impact the critical path and add a negligible area overhead of \SI{0.13}{\percent}.

The remainder of this paper is structured as follows: \autoref{s:background} presents our threat model, the attack algorithm and the concept of \emph{time protection} as a prevention methodology.
We derive two security requirements that hardware needs to provide to enable time protection in \autoref{s:sec-req}.

In \autoref{s:impl}, we propose different implementation approaches for time protection, which we evaluate in \autoref{s:eval}.
We present the work related to this paper in \autoref{s:related-work} and conclude in \autoref{s:conclusions}.

\section{Background}
\label{s:background}

\subsection{Timing Channels}

Information channels that let isolated applications communicate but are not intended for information transfer are called \emph{covert channels}~\cite{Lampson_73}.
\emph{Microarchitectural timing channels} are covert channels that leverage the timing behaviour of microarchitectural components to transfer information.
They may generally occur whenever multiple applications compete for shared hardware resources~\cite{Ge2018Survey}.
Famous examples for exploitable hardware resources are data caches, where the latency of a memory access varies depending on how the cache was previously used~\cite{Hu_92,percival2005cache}.
Attacks were also demonstrated for further components such as instruction caches~\cite{Aciiccmez2007ICache}, branch predictors~\cite{Aciiccmez2007BP}, and \glspl{tlb}~\cite{Gras2018TLB}.

\subsection{Threat Model}

\begin{figure}
    \centering
    \includegraphics[width=0.95\columnwidth]{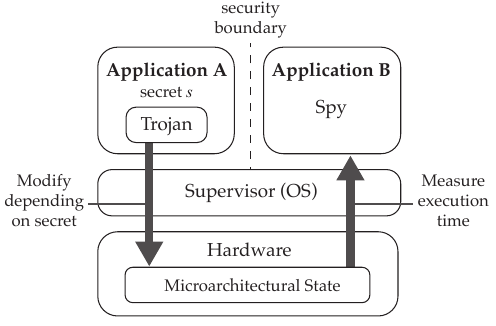}
    \caption{Threat model: a Trojan actively leaking data via shared hardware resources.}
    \label{f:cc}
\end{figure}

We examine covert-channel leakage under a \emph{confinement} scenario~\cite{Lampson_73}: An untrusted program possesses a secret, and the \gls{os} encapsulates the program's execution in a security domain that only allows communication across defined channels to trusted components (e.g., an encryption service).
The untrusted program contains a Trojan that is actively trying to leak the secret via a covert channel.
Note that a Trojan could not only hide in malicious code, but can be constructed by control-flow hijacking of innocent code through exploiting bugs or speculatively executed gadgets as in a Spectre attack~\cite{Kocher2018spectre}.
A second, unconfined, and also untrusted security domain contains a spy which is trying to read the secret leaked by the Trojan.
This setup is illustrated in \autoref{f:cc}.

The \emph{intentional} leakage by the Trojan represents the worst case. If we can prevent this attack, we preclude any other leakage using the same mechanism including \emph{side channels}, where leakage originates from an unwitting victim rather than a Trojan.

We assume that the Trojan and spy time-share one processor core, meaning cross-core leakage is out of the scope of this paper.
\footnote{The prevention of cross-core leakage is discussed further by Ge et al.~\cite{Ge2019a}.}
We only consider microarchitectural timing channels.
Covert channels that abuse other characteristics, such as power draw, are not covered in this work.

\subsection{Time Protection}
\label{s:time-protection}

Time protection is a principled approach to  \emph{preventing} timing channels~\cite{Ge2019a}.
While the established notion of \emph{memory protection} prevents interference between security domains through unauthorised memory accesses, time protection aims to prevent interference that affects observable timing behaviour.

Time protection requires that all shared hardware resources, including non-architectural ones, must be partitioned between security domains, either temporally (secure time multiplexing) or spatially.
Ge et al. show that (physically-addressed) off-core caches can be effectively partitioned through \emph{cache colouring}~\cite{kessler1992colouring}, which leverages the associative cache lookup to force different partitions into disjoint subsets of the cache.
They demonstrate that colouring is effective in preventing cache channels in both intra-core and cross-core attacks and comes with low overhead.

Spatial partitioning is generally impractical for on-core resources.
For performance reasons, on-core resources are limited and are designed to be well utilised by a single program, so partitioning approaches usually result in unacceptable performance degradation.
Furthermore, on-core resources are generally indexed by virtual addresses, which cannot be coloured by the \gls{os}.
This leaves temporal partitioning as the only viable approach for on-core resources.

\section{Security Requirements}
\label{s:sec-req}

Based on the findings of Ge et al.~\cite{Ge2019a}, we propose two security requirements for on-core time protection by temporal partitioning.

First, before handing a resource to a different domain, it must be brought to a state that is independent of execution history.
Therefore, the \gls{os} must have the means to clear all microarchitectural state, which in practical terms requires an extension to the hardware-software contract to refer (in an abstract way) to such non-architectural state.
Ge et at.~specifically show that contemporary Intel and Arm processors lack the mechanisms required for implementing time protection~\cite{Ge2018}.

\begin{requirement}
    \label{r:flush}
    For temporal partitioning, hardware must provide a (set of) mechanism(s) that allow clearing all non-architectural state that depends on previous execution and may impact future timing.
\end{requirement}

A second requirement for time protection is a secret-independent context switch latency.
This requirement is particularly (but not exclusively) relevant for processors featuring a write-back L1 cache: before we can reset this component, we need to write back all dirty cache lines.
The number of dirty cache lines, and hence the latency of said reset and the whole context switch routine, directly depends on previous execution and can be probed to transfer information.

\begin{requirement}
    \label{r:pad}
    The latency of the context switch routine, including the reset of the non-architectural state mentioned above, needs to be independent of the previous execution.
\end{requirement}

\section{Implementing On-Core Time Protection}
\label{s:impl}

In the following, we will present different approaches to implement time protection for shared on-core hardware resources.
We will use the \riscv{} architecture for our analysis because of its openness, extensibility, and availability of modifiable open-source implementations such as CVA6 (see \autoref{s:hwplatform}).
\autoref{s:sw} will describe a baseline system where time protection will be implemented using existing resources on unmodified hardware without additional architectural features.
In \autoref{s:fencet}, we will propose an \gls{isa} extension that provides software with the necessary means to partition shared on-core hardware resources temporally (\autoref{r:flush} of on-core time protection), and we will discuss three different implementation approaches.
Finally, \autoref{s:pad} will address \autoref{r:pad} of time protection, enabling a context-switch latency that does not depend on execution history.

\subsection{Baseline Architecture}
\label{s:sw}

Ge et al. \cite{Ge2019a} report that neither the x86 nor the Arm architecture provides sufficient mechanisms for implementing time protection.
Arm provides targeted L1 cache flushes but no mechanism for flushing other microarchitectural state.
The x86 architecture provides \emph{branch control mechanisms} for clearing the state of the branch predictors~\cite{Intel2018IBC}.
For partitioning the L1 cache on this architecture, the authors implemented software flushing by touching all cache lines, similar to the prime phase of the prime-and-probe attack.
Such an approach is expensive and obviously brittle, as it must make assumptions on the replacement policy which may not hold in reality.
Unsurprisingly, they find that this defence is incomplete, leaving residual channels that the \gls{os} is unable to close.

With \riscv{}, the situation is presently worse, as the specification of cache management is still under discussion.
While implementations generally support some cache management, this is not yet standardised.
To explore this aspect, we implement a ``software only'' defence (in the following referred to as \emph{Software}, SW), where the \gls{os} uses only mechanisms defined in the \gls{isa} as presently specified.
This basically forces the \gls{os} to resort to the priming approach in an attempt to erase any microarchitectural state left by the Trojan's execution.

\subsection{Temporal Fence Instruction}
\label{s:fencet}

As we show in \autoref{s:eval}, this ``software only'' defence is insufficient since some microarchitectural timing channels remain.
Moreover, it comes at a great performance overhead.
Therefore, we propose the \emph{temporal fence} instruction, \fencet{}, to let an \gls{os} access the hardware mechanisms required for time protection.
We note that other semantics to realise the temporal fence are conceivable as well: for instance, a combination of multiple instructions and registers could be used.
For our proof of concept, we stick to a single \fencet{} instruction that both triggers the reset of vulnerable microarchitectural state (\autoref{r:flush}) and guarantees a history-independent context switch latency (\autoref{r:pad}).
Except for increased cycle and instruction counters, \fencet{} has no architectural effects.
For evaluation purposes, we encode \fencet{} as a U-type \riscv{} instruction with the opcode \emph{custom-0}.
\fencet{} optionally takes a 20-bit immediate value, which is a bitmap selecting the components that should be reset.

In the following, we will present three different implementation strategies for \fencet{}.

\subsubsection{Basic Flush}
\label{s:selfirst}

In the first version of \fencet{}, we exclusively flush the principal components used by the prime-and-probe attack in \autoref{s:prime-probe}: the L1 data and instruction caches, the \glspl{tlb}, and the branch predictors (\gls{bht}, \gls{btb}).
If present, prefetchers are cleared and write-buffers are drained.
We write back any dirty state (for write-back caches), invalidate the caches and \glspl{tlb}, and purge the branch predictors.
To preserve the computational correctness of in-flight instructions, we also flush the pipeline.

In the following, we will refer to this approach as the \emph{Basic Flush} (\firstflush{}).

\subsubsection{Full Flush}
\label{s:selall}

Motivated by our security analysis of the basic flush in \autoref{s:eval-sec}, we identify several secondary, stateful components deeply embedded in CVA6 with a possible timing impact.
In particular, these are
\begin{itemize}
    \item a \gls{lfsr} per cache (L1 data, L1 instruction) that generates a pseudo-random number sequence for the cache-replacement policy,
    \item a \gls{plru} tree in each \gls{tlb} that identifies a replacement candidate,
    \item a round-robin memory arbiter that arbitrates cache accesses between the load unit, the store unit, and the memory-management unit,
    \item two round-robin arbiters in the write buffer of the write-through L1 data cache, which choose an entry to serve (lookup or write back) next.
\end{itemize}
We extend \fencet{} to a \emph{full flush} (\fullflush{}) by adding the support to clear the state of these components as well.

\subsubsection{Microreset}
\label{s:urst}

Finally, we propose a principled and systematic approach to enforce complete temporal partitioning.
The key idea is to clear \emph{all} on-core state that is not architectural by default, and explicitly exclude architectural state.
We call this mechanism \textsc{Microreset}, as it exclusively resets non-architectural microarchitectural state.

All flip-flops in the design are extended by an additional \texttt{clear} input.
By asserting this input on a \fencet{}, we can guarantee that all state on flip-flops in the design is set back to a predefined state.
On-core state that is not resettable (such as SRAMs) must be cleared separately.
To ensure computational correctness, architectural state needs to be retained, either by saving it before \textsc{Microreset} (e.g.\ write-back of an L1 cache) or by explicitly excluding it from the \textsc{Microreset}.
Hence, we design the \fencet{} controller to proceed in the following six steps:

\begin{step}
    Save the program counter.
\end{step}

To resume execution after \textsc{Microreset} from the correct location, we store the address of the instruction following \fencet{} in a register that is excluded from  \textsc{Microreset} (we consider this architectural state).

\begin{step}
    Save locally modified (dirty) architectural state.
\end{step}

In particular, this concerns components such as write-back L1 caches: to preserve the contents of dirty cache lines, we need to write them back before clearing the cache.
As we will discuss later, this is the most costly step of \fencet{}.

\begin{step}
    Drain pending transactions.
\end{step}

Next, we need to wait for all pending external transactions to complete without issuing or accepting new ones.
This way, we prevent violating any handshake protocols or losing data that we began to write back in the previous step.

\begin{step}
    Clear components that are not cleared on reset.
\end{step}

Not all components can be fully reset to a predefined state.
For instance, SRAMs such as those found in caches are generally not resettable.
To temporally partition these components, they need to be cleared separately, e.g.\ by a \gls{fsm} that overwrites their contents line by line.

\begin{figure}
    \centering
    \includegraphics[width=0.8\columnwidth]{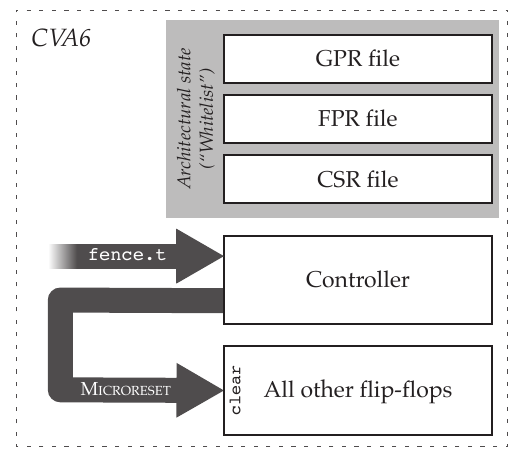}
    \caption{Illustration of the \textsc{Microreset}.}
    \label{f:urst}
\end{figure}

\newpage

\begin{step}
    Assert \textsc{Microreset}.
\end{step}

We now clear all flip-flops containing non-architectural state.
For this purpose, we assert the \texttt{clear} input of all flip-flops in the design.
We explicitly exclude flip-flops that hold architectural state---the state of these components is preserved during \textsc{Microreset} and saved/restored explicitly by the \gls{os} during a context switch.
This approach removes the risk of omitting state that could create a timing channel.

While identifying the architectural state is potentially one of the biggest challenges of this approach, for CVA6, it turned out relatively straight-forward: we explicitly exclude the integer and floating-point register files, the \gls{csr} file, and the controller, which is driving \textsc{Microreset}.
\autoref{f:urst} shows the resulting setup.

In other designs, such as out-of-order cores with merged register files, identifying the architectural state might be more challenging but should, in general, still be feasible with reasonable effort.

\begin{step}
    Continue execution from saved program counter.
\end{step}

Finally, we de-assert the \textsc{Microreset} and continue fetching from the next program counter.

\subsection{Time Padding}
\label{s:pad}

When introducing time protection in \autoref{s:sec-req}, we asserted that the context switch latency needs to be independent of previous execution (\autoref{r:pad}).

In \riscv{}, the context switch routine is initiated by a timer interrupt of the \gls{clint}.
While this interrupt is generated at a fixed period independently of the microarchitectural state, any machine-mode or kernel code following it until the end of \fencet{} may be delayed by cache misses, mispredictions etc.
Hence, to prevent a dependency of the context switch latency on previous execution, we pad the interval between the \gls{clint}'s timer interrupt and the completion of \fencet{} to a worst-case latency.

We implement this mechanism by adding a custom \cspad{} \gls{csr} that takes a 32-bit value.
We stall the completion of \fencet{} until \cspad{} cycles after the \gls{clint} timer interrupt.
Padding can be disabled by setting \cspad{} to 0.

\section{Evaluation}
\label{s:eval}

\subsection{Hardware Platform}
\label{s:hwplatform}

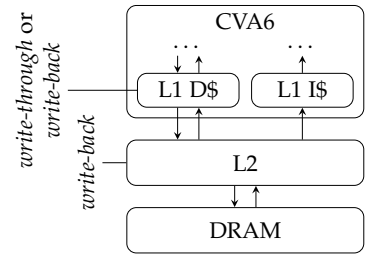
\begin{figure}[t]
    \centering
    \scalebox{0.9}{\def\NodeHeight{1cm}
\def\NodeHeightTwo{2em}
\def\InnerNodeHeight{0.5cm}
\def\NodeWidth{3.5cm}
\def\InnerNodeWidth{1.5cm}
\def\NodeDist{2ex}
\def\edgedist{2cm}
\tikzstyle{Box} = [draw, rounded corners, minimum height=\NodeHeight, minimum width=\NodeWidth, node distance=\NodeDist, align=center]
\tikzstyle{BoxTwo} = [draw, rounded corners, minimum height=\NodeHeightTwo, minimum width=\NodeWidth, node distance=\NodeDist, align=center]
\tikzstyle{InsideBox} = [draw, rounded corners, minimum height=\InnerNodeHeight, minimum width=\InnerNodeWidth, align=center]
\tikzstyle{Desc} = [node distance=0.3cm, align=right]

\begin{tikzpicture}[x=1ex,y=1ex]
  
		\node (ariane) [BoxTwo, text depth=1.2cm] {CVA6};
		\node[InsideBox, anchor=south west] (l1d) at ($(ariane.south west)+(1,1)$) {L1 D\$};
		\node[InsideBox, anchor=south east] (l1i)    at ($(ariane.south east)+(-1,1)$) {L1 I\$};
		\node (l1ddots) at ($(l1d.north)+(0,2.5)$) {\ldots};
		\node (l1idots) at ($(l1i.north)+(0,2.5)$) {\ldots};
		\node (l2) 	  [BoxTwo, below=of ariane] {L2};
		\node (dram)	  [BoxTwo, below=of l2]      {DRAM};
		\node[anchor=south, rotate=90, align=center] (wt) at ($(l1d.west)+(-6.5,0)$) {\textit{write-through} or \\ \textit{write-back}};
		\node[anchor=north, rotate=90] (wb) at ($(wt.center|-l2.west)+(3,0)$) {\textit{write-back}};
	
		\path (l1d.south)+(-1,0) coordinate (l1dsource);
		\path (l1d.south)+(1,0)  coordinate (l1dsink);
	
		\draw [-stealth] (l1i.north) -- (l1idots.south);
		\draw [-stealth] ($(l1d.north)+(1,0)$) -- ($(l1ddots.south)+(1,0)$);
		\draw [-stealth] ($(l1ddots.south)-(1,0)$) -- ($(l1d.north)-(1,0)$);
		\draw [-stealth] (l1dsource) -- (l1dsource|-l2.north);
		\draw [-stealth] (l1dsink|-l2.north) -- (l1dsink);
		\draw [-stealth] (l1i.south|-l2.north) -- (l1i.south);
		\draw [-stealth] ($(dram.north)+(1,0)$) -- ($(l2.south)+(1,0)$);
		\draw [-stealth] ($(l2.south)-(1,0)$) -- ($(dram.north)-(1,0)$);
		\draw [-] (wt.south|-l1d.west) -- (l1d.west);
		\draw [-] (wb.south) -- (l2.west);
\end{tikzpicture}}
    \caption{Hardware platform.}
    \label{fig:hw-platform}
\end{figure}

We evaluate the channels and defences on CVA6, an open-source, RV64GC, 6-stage RISC-V core developed at ETH Z\"urich and currently maintained by OpenHW Group~\cite{Zaruba2019}.\footnote{CVA6 is formerly known as \textsc{Ariane}.}
It is implemented in SystemVerilog and publicly available on GitHub~\cite{GitHub:Ariane}.
It features three privilege levels and address translation, and thus supports full-fledged operating systems.
Its configurability, simplicity, and openness make it a good candidate for architectural exploration.

\paragraph*{Setup}
We instantiate the CVA6 core on a Xilinx Kintex-7 FPGA (Digilent Genesys II), running at \SI{50}{\mega\hertz}.
We configure two versions of CVA6: one with a \emph{write-through} L1 data cache and one with a \emph{write-back} cache.
Both versions feature an 8-way, 32\,KiB L1 data cache and a 4-way, 16\,KiB L1 instruction cache.
The caches use 16-byte lines and a pseudo-random replacement strategy driven by an 8-bit \gls{lfsr}.
The L1 data cache is accessed by the \mbox{load-,} \mbox{store-,} and memory-management units, with concurrent accesses arbitrated with a round-robin policy.
The branch predictor has a 64-entry \gls{bht} and a 16-entry \gls{btb}.
There are two single-level, fully associative data and instruction \glspl{tlb}, with 16 entries each, using a \gls{plru} replacement policy.
Our system-on-chip features a 512-KiB write-back L2 cache~\cite{wolfgang2019llc-thesis} that is connected to DRAM.
\autoref{fig:hw-platform} shows the memory architecture.

We partition the L2 cache by colouring~\cite{kessler1992colouring}, which precludes channels in the memory backend and allows us to focus on channels resulting from on-core state.

\subsection{Security Analysis}
\label{s:eval-sec}

\subsubsection{Prime and Probe}
\label{s:prime-probe}

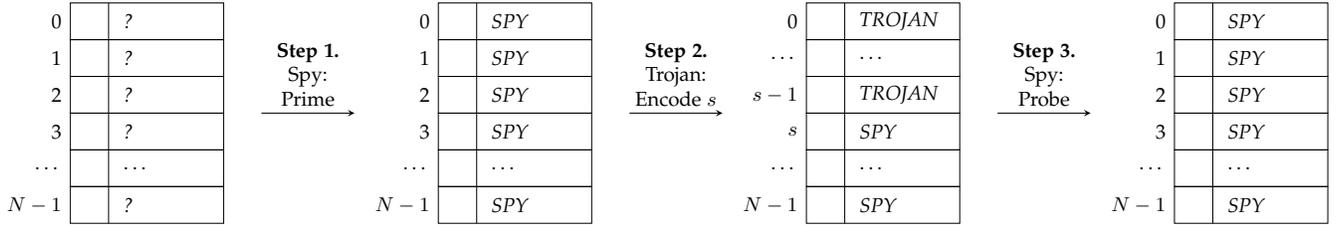
\begin{figure*}
    \centering
    \scalebox{0.8}{\begin{tikzpicture}[x=0.14\linewidth,y=1.5\baselineskip]

  \begin{scope}[local bounding box=state1]

    \foreach \x in {0,1,...,5}
    {
      \node[inner sep=1pt] (BL\x) at (0,-\x)   {};
      \node[inner sep=1pt] (UR\x) at (1,-\x+1) {};
      \node (MM\x) at ($(BL\x)!0.5!(UR\x)$) {};
      \draw (BL\x) rectangle (UR\x);
    }

    \node[anchor=west] at ($(MM0)-(0.2,0)$)  {\itshape ?};
    \node[anchor=east] at ($(MM0)-(0.51,0)$) {0};
    \node[anchor=west] at ($(MM1)-(0.2,0)$)  {\itshape ?};
    \node[anchor=east] at ($(MM1)-(0.51,0)$) {1};
    \node[anchor=west] at ($(MM2)-(0.2,0)$)  {\itshape ?};
    \node[anchor=east] at ($(MM2)-(0.51,0)$) {2};
    \node[anchor=west] at ($(MM3)-(0.2,0)$)  {\itshape ?};
    \node[anchor=east] at ($(MM3)-(0.51,0)$) {3};
    \node[anchor=west] at ($(MM4)-(0.2,0)$)  {\ldots};
    \node[anchor=east] at ($(MM4)-(0.51,0)$) {\ldots};
    \node[anchor=west] at ($(MM5)-(0.2,0)$)  {\itshape ?};
    \node[anchor=east] (nsets) at ($(MM5)-(0.51,0)$) {$N - 1$};

    \draw ($(BL0|-UR0)+(0.25,0)$) -- ($(BL5)+(0.25,0)$);

    \node (state1right) at ($(UR3)+(0.2,0)$) {};
  \end{scope}

  \begin{scope}[shift={(2.4,0)}, local bounding box=state1]

    \foreach \x in {0,1,...,5}
    {
      \node[inner sep=1pt] (BL\x) at (0,-\x)   {};
      \node[inner sep=1pt] (UR\x) at (1,-\x+1) {};
      \node (MM\x) at ($(BL\x)!0.5!(UR\x)$) {};
      \draw (BL\x) rectangle (UR\x);
    }

    \node[anchor=west] at ($(MM0)-(0.2,0)$)  {\itshape SPY};
    \node[anchor=east] at ($(MM0)-(0.51,0)$) {0};
    \node[anchor=west] at ($(MM1)-(0.2,0)$)  {\itshape SPY};
    \node[anchor=east] at ($(MM1)-(0.51,0)$) {1};
    \node[anchor=west] at ($(MM2)-(0.2,0)$)  {\itshape SPY};
    \node[anchor=east] at ($(MM2)-(0.51,0)$) {2};
    \node[anchor=west] at ($(MM3)-(0.2,0)$)  {\itshape SPY};
    \node[anchor=east] at ($(MM3)-(0.51,0)$) {3};
    \node[anchor=west] at ($(MM4)-(0.2,0)$)  {\ldots};
    \node[anchor=east] at ($(MM4)-(0.51,0)$) {\ldots};
    \node[anchor=west] at ($(MM5)-(0.2,0)$)  {\itshape SPY};
    \node[anchor=east] (nsets) at ($(MM5)-(0.51,0)$) {$N - 1$};

    \draw ($(BL0|-UR0)+(0.25,0)$) -- ($(BL5)+(0.25,0)$);

    \node (state2right) at ($(UR3)+(0.2,0)$) {};
    \node (state2left)  at ($(BL2)-(0.5,0)$) {};
  \end{scope}

  \begin{scope}[shift={(4.8,0)}, local bounding box=state2]

    \foreach \x in {0,1,...,5}
    {
      \node[inner sep=1pt] (BL\x) at (0,-\x)   {};
      \node[inner sep=1pt] (UR\x) at (1,-\x+1) {};
      \node (MM\x) at ($(BL\x)!0.5!(UR\x)$) {};
      \draw (BL\x) rectangle (UR\x);
    }

    \node[anchor=west] at ($(MM0)-(0.2,0)$)  {\itshape TROJAN};
    \node[anchor=east] at ($(MM0)-(0.51,0)$) {0};
    \node[anchor=west] at ($(MM1)-(0.2,0)$)  {\ldots};
    \node[anchor=east] at ($(MM1)-(0.51,0)$) {\ldots};
    \node[anchor=west] at ($(MM2)-(0.2,0)$)  {\itshape TROJAN};
    \node[anchor=east] at ($(MM2)-(0.51,0)$) {$s-1$};
    \node[anchor=west] at ($(MM3)-(0.2,0)$)  {\itshape SPY};
    \node[anchor=east] at ($(MM3)-(0.51,0)$) {$s$};
    \node[anchor=west] at ($(MM4)-(0.2,0)$)  {\ldots};
    \node[anchor=east] at ($(MM4)-(0.51,0)$) {\ldots};
    \node[anchor=west] at ($(MM5)-(0.2,0)$)  {\itshape SPY};
    \node[anchor=east] (nsets) at ($(MM5)-(0.51,0)$) {$N - 1$};

    \draw ($(BL0|-UR0)+(0.25,0)$) -- ($(BL5)+(0.25,0)$);

    \node (state3right) at ($(UR3)+(0.2,0)$) {};
    \node (state3left)  at ($(BL2)-(0.5,0)$) {};
  \end{scope}

  \begin{scope}[shift={(7.2,0)}, local bounding box=state2]

    \foreach \x in {0,1,...,5}
    {
      \node[inner sep=1pt] (BL\x) at (0,-\x)   {};
      \node[inner sep=1pt] (UR\x) at (1,-\x+1) {};
      \node (MM\x) at ($(BL\x)!0.5!(UR\x)$) {};
      \draw (BL\x) rectangle (UR\x);
    }

    \node[anchor=west] at ($(MM0)-(0.2,0)$)  {\itshape SPY};
    \node[anchor=east] at ($(MM0)-(0.51,0)$) {0};
    \node[anchor=west] at ($(MM1)-(0.2,0)$)  {\itshape SPY};
    \node[anchor=east] at ($(MM1)-(0.51,0)$) {1};
    \node[anchor=west] at ($(MM2)-(0.2,0)$)  {\itshape SPY};
    \node[anchor=east] at ($(MM2)-(0.51,0)$) {2};
    \node[anchor=west] at ($(MM3)-(0.2,0)$)  {\itshape SPY};
    \node[anchor=east] at ($(MM3)-(0.51,0)$) {3};
    \node[anchor=west] at ($(MM4)-(0.2,0)$)  {\ldots};
    \node[anchor=east] at ($(MM4)-(0.51,0)$) {\ldots};
    \node[anchor=west] at ($(MM5)-(0.2,0)$)  {\itshape SPY};
    \node[anchor=east] (nsets) at ($(MM5)-(0.51,0)$) {$N - 1$};

    \draw ($(BL0|-UR0)+(0.25,0)$) -- ($(BL5)+(0.25,0)$);

    \node (state4left) at ($(BL2)-(0.5,0)$) {};
  \end{scope}

  \draw [-stealth] (state1right) -- node[above,align=center] {\textbf{Step 1.}\\Spy:\\Prime} (state2left);
  \draw [-stealth] (state2right) -- node[above,align=center] {\textbf{Step 2.}\\Trojan:\\Encode $s$} (state3left);
  \draw [-stealth] (state3right) -- node[above,align=center] {\textbf{Step 3.}\\Spy:\\Probe} (state4left);

\end{tikzpicture}}
    \caption{A prime-and-probe attack on a cache.}
    \label{fig:prime}
\end{figure*}

Techniques for exploiting covert channels are well established; for our scenario of intentional leakage, the \emph{prime-and-probe} attack~\cite{percival2005cache} is simple and effective.
We stress that our proposed mechanism addresses the root cause of covert channels and therefore expect an equal efficacy for other attacks such as \emph{evict-and-time}~\cite{Osvik2006EvictAndTime} and \emph{evict-and-reload}~\cite{Gruss2015EvictAndReload}.

In a prime-and-probe attack, the spy first forces the exploited hardware resource into a known state (\emph{prime}).
For the data cache it traverses a large buffer (in cache-line-sized strides for efficiency); for the instruction cache it executes a series of linked jumps.
The \glspl{tlb} are similarly primed by accessing or jumping with page-size strides.\footnote{This is a somewhat simplified description---in general, it is necessary to randomise the access order to prevent interference from prefetching, but that is not an issue on our processor.}
The branch predictors are primed by a series of conditional branches (\gls{bht}) or by executing multiple indirect jumps (\gls{btb}).
With a correctly-sized priming buffer, this leaves the hardware resources in a state where further accesses by the spy within the same address range are fast. This state is illustrated in the second element of \autoref{fig:prime} where all entries in the buffer have been primed by the spy.
At the end of its time slice, the \gls{os} preempts the spy and switches to the application that contains the Trojan, which accesses a subset of the hardware resource to encode the secret.
Given a cache of \(n\) lines, the Trojan can transmit a secret \(s\leq n\), the \emph{input signal}, by touching \(s\) cache lines, thereby replacing the spy's content.
The resulting state is illustrated in the third element of \autoref{fig:prime}.
Obviously, more complex encodings are possible to increase the amount of data transferred in a time slice (the channel capacity), but for our purposes, the simple encoding is sufficient, as we want to prevent \emph{any} leakage.

When execution switches back to the spy, it again traverses (\emph{probes}) the whole buffer, observing its execution time.
Each entry replaced by the Trojan's execution leads to a cache miss, and results in an increase in probe time.
If the latency of a hit is \(t_{\textrm{hit}}\) and that of a miss is \(t_{\textrm{miss}} > t_{\textrm{hit}}\), the total latency increase is \(s \cdot (t_{\textrm{miss}} - t_{\textrm{hit}}).\)
For our simple encoding scheme, the \emph{output signal} is the total probe time, which is linearly correlated to the input signal.
A more sophisticated encoding scheme could, for example, exploit the time measurements of each individual access and thus extract more information.

\subsubsection{Measuring Leakage}

We adopt the approach of Ge et al. \cite{Ge2018} for quantifying and evaluating leakage and prevention strategies.
For attack \(i\), the Trojan encodes as input value a randomly chosen secret, \(s_i\), and the spy subsequently measures as the output value its probe latency, \(t_i\).
$s$ and $t$ can be regarded as samples of the random variables $S$ and $T$.
A covert channel exploits the correlation of the two random variables: if the output $t$ is correlated with the input $s$, there is a covert channel that transfers information from the Trojan to the spy.

We use a \emph{sample size} (number of repeated attacks) of 1~million.
For leakage we use a combination of two indicators: The \emph{channel matrix} for visualisation and the \emph{discrete mutual information} $\mathcal{M}$ as a quantitative metric.

\paragraph{Channel Matrix}

The channel matrix represents the conditional probability of observing a particular output value, $t$, given input value $s$.
The conditional probability distribution ${p(t\mid s)}$ can be computed directly from the measured sample pairs $\{(s_1, t_1), \ldots , (s_N, t_N)\}$.

We represent the channel matrix as a heat map: inputs vary horizontally and outputs vertically, and bright colours indicate high, dark colours low probability.
A variation of colour along any horizontal line through the graph indicates a dependence of the output on the input, and thus a channel.
For example, \autoref{f:cm-base} shows a clear diagonal pattern indicating a channel: if the spy observes a probe time of 85,000 cycles, it can infer with high confidence that the Trojan has encoded a value between \emph{170} and \emph{180}.
Repeating the experiment can increase the spy's confidence.

\paragraph{Mutual Information}

For quantifying channel capacity we use \emph{continuous mutual information} $\mathcal{M}$, the amount of information gained about a random variable by observing another, possibly correlated random variable~\cite{Shannon_48}.

Intuitively, mutual information is the difference of the information gained by observing the random variable $T$ \emph{without} and \emph{with} knowledge of the second random variable $S$.
If both random variables are highly correlated (i.e., there exists a covert channel), the information gained by observing $S$ is low and $\mathcal{M}$ high.
Conversely, if both random variables are uncorrelated, $\mathcal{M} = 0$.
The unit for $\mathcal{M}$ is bits; as most of our channel capacities are small, we use millibits ($1 \mathrm{\,mb}=10^{-3}\,\mathrm{b}$) in our measurements.

\paragraph{Zero Leakage Upper Bound $\mathcal{M}_0$}
\label{s:m0}
Since all measurements are affected by noise, $\mathcal{M}$ will mostly not be zero, even if there is no channel.
We use a Monte Carlo simulation for estimating the apparent channel produced by this noise.
Specifically, we pick uniformly random pairs of input and output values, and thus remove any correlation between them, while retaining their original value ranges and spreads.
Any mutual information that is measured from this data can only be due to noise.
We repeat this process 1000 times and then compute the 95\%-confidence interval $\mathcal{M}_0$ for an experiment without a channel.
Notably, $\mathcal{M}_0$ can differ strongly between experiments, as it depends on the range and distribution of the measured values. 
We conclude that a channel is present if $\mathcal{M} > \mathcal{M}_0$, otherwise, the result is consistent with no channel.

We use the leakiEst tool~\cite{Chothia2013} to compute mutual information $\mathcal{M}$ and zero leakage upper bounds $\mathcal{M}_0$.

\paragraph{Testbench}
\label{s:testbench}

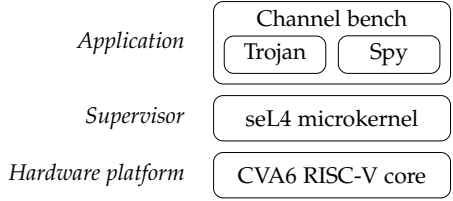
\begin{figure}[t]
    \centering
    \scalebox{0.9}{\def\NodeHeight{1cm}
\def\NodeHeightTwo{2em}
\def\InnerNodeHeight{0.5cm}
\def\NodeWidth{3.5cm}
\def\InnerNodeWidth{1.5cm}
\def\NodeDist{1ex}
\def\edgedist{2cm}
\tikzstyle{Box} = [draw, rounded corners, minimum height=\NodeHeight, minimum width=\NodeWidth, node distance=\NodeDist, align=center]
\tikzstyle{BoxTwo} = [draw, rounded corners, minimum height=\NodeHeightTwo, minimum width=\NodeWidth, node distance=\NodeDist, align=center]
\tikzstyle{InsideBox} = [draw, rounded corners, minimum height=\InnerNodeHeight, minimum width=\InnerNodeWidth, align=center]
\tikzstyle{Desc} = [node distance=0.3cm, align=right]

\begin{tikzpicture}[x=1ex,y=1ex]
	\node (ChanBench) [Box, text depth=0.75\NodeHeight] {Channel bench};
	\node[InsideBox, anchor=south west] (Trojan) at ($(ChanBench.south west)+(1,1)$) {Trojan};
	\node[InsideBox, anchor=south east] (Spy)    at ($(ChanBench.south east)+(-1,1)$) {Spy};
	\node (sel4) 	  [BoxTwo, below=of ChanBench] {seL4 microkernel};
	\node (Ariane)	  [BoxTwo, below=of sel4]      {CVA6 RISC-V core};
	
	\node (Application) [Desc, left=of ChanBench] {\itshape Application};
	\node (Supervisor)	[Desc, left=of sel4]      {\itshape Supervisor};
	\node (HWPlat)		[Desc, left=of Ariane]	  {\itshape Hardware platform};
	
	\path (Ariane.south east)+(2,0) node(start) {};
	\path (ChanBench.north east)+(2,0) node(end) {};
	
\end{tikzpicture}}
    \caption{HW/SW stack of the evaluation framework.}
    \label{f:stack}
\end{figure}

Ge's \textsc{Channel Bench}~\cite{Ge2019phd,GitHub:channel-bench} provides a minimal \gls{os} and data collection infrastructure; we port it to \riscv{} and adapt it to CVA6.
\textsc{Channel Bench} uses attack implementations from the \textsc{Mastik} toolkit~\cite{Yarom2016}, running on an experimental version of seL4~\cite{Klein2014_seL4} that supports time protection.
The resulting stack of our evaluation framework is shown in \autoref{f:stack}.

\subsubsection{L1 Data Cache}
\label{s:channels-l1}

\newlength\figH
\newlength\figW
\setlength{\figH}{0.6\linewidth} %
\setlength{\figW}{1.618\figH}
\def\cmScale{0.73}

\pgfplotsset{cmStyle/.style={
    x tick label style={/pgf/number format/.cd, precision=1},
    scaled y ticks = false,
    y tick label style={/pgf/number format/.cd, fixed, precision=0},
    xtick distance = 64,
    colorbar style = {at={(1.1,0)}, anchor=south west, scaled y ticks=false}
}}

\begin{figure}
    \centering
    \begin{subfigure}[t]{\linewidth}
        \centering
        \begin{tikzpicture}[scale=\cmScale]

\definecolor{color0}{rgb}{0.267004,0.004874,0.329415}

\begin{axis}[cmStyle,
cmStyle,
axis background/.style={fill=color0},
colorbar,
colorbar style={ytick={0,0.00964909860909337,0.013876876089505,0.0163499673769393,0.0181046535699167,0.0194656939197933,0.0205777448573509,0.0215179705280914,0.0223324310503284,0.0230508361447851,0.023693471400205,0.0279212488806167,0.0303943401680509},yticklabels={\(\displaystyle {0}\),\(\displaystyle {10^{-3}}\),,,,,,,,,\(\displaystyle {10^{-2}}\),,},ylabel={Probability}},
colormap/viridis,
height=\figH,
point meta max=0.03049394860864,
point meta min=0,
tick align=outside,
tick pos=left,
width=\figW,
x grid style={white!69.0196078431373!black},
xlabel={Secret},
xmin=0, xmax=257,
xtick style={color=black},
y grid style={white!69.0196078431373!black},
ylabel={Time (cycles)},
ymin=76254, ymax=89564,
ytick={77500,80000,82500,85000,87500},
ytick style={color=black}
]
\addplot graphics [includegraphics cmd=\pgfimage,xmin=0, xmax=257, ymin=76254, ymax=89564] {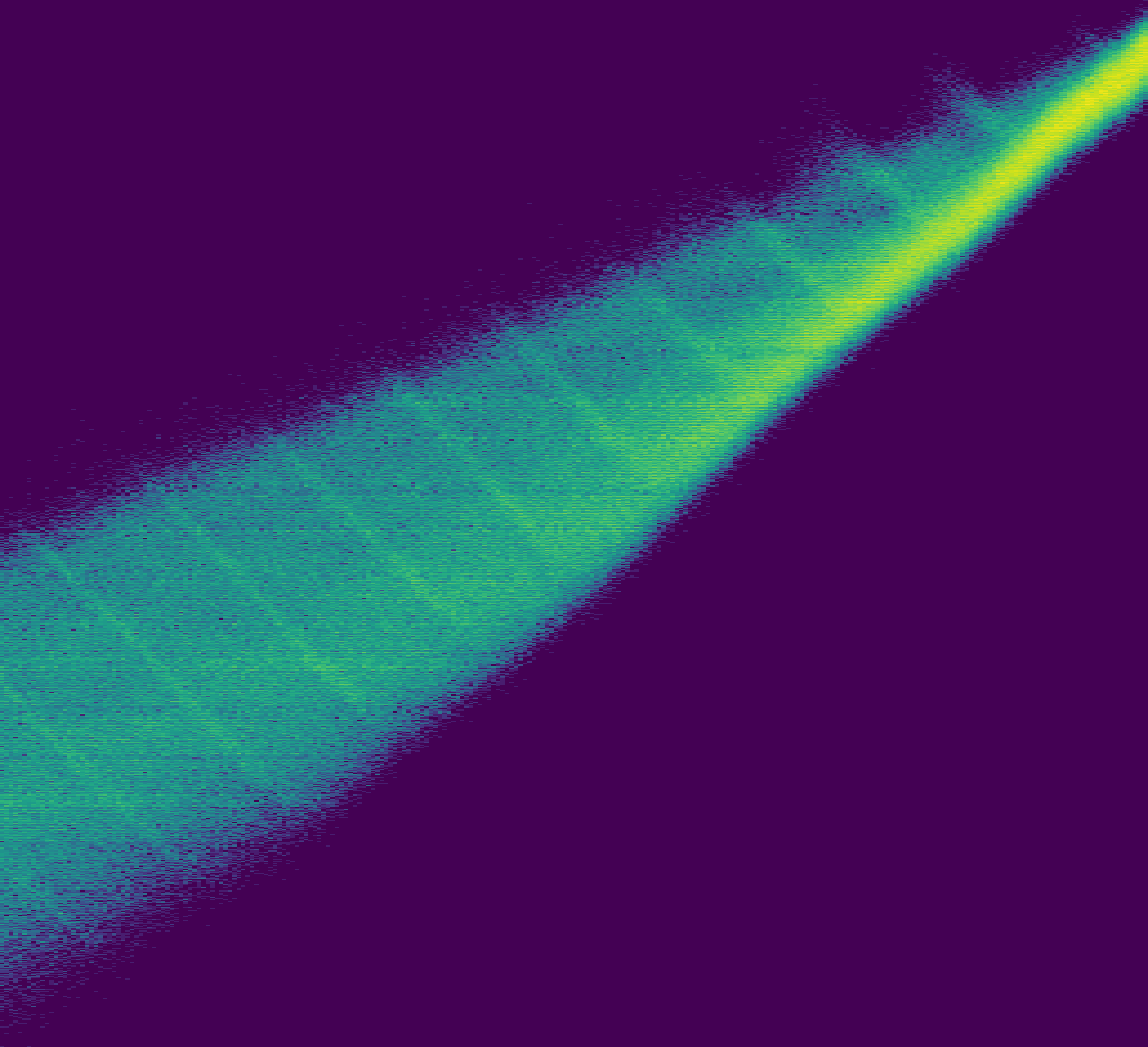};
\draw[->,very thick,draw=red] (0, 85000) -- (172, 85000);
\draw[->,very thick,draw=red] (176, 85000) -- (176, 76250);
\end{axis}

\end{tikzpicture}
        \caption{Unmitigated. (1629/0.5)}
        \label{f:cm-base}
    \end{subfigure}
    \\\vspace{2em}
    \begin{subfigure}[t]{\linewidth}
        \centering
        \begin{tikzpicture}[scale=\cmScale]

\definecolor{color0}{rgb}{0.267004,0.004874,0.329415}

\begin{axis}[cmStyle,
axis background/.style={fill=color0},
colorbar,
colorbar style={ytick={0,0.00969801765177042,0.0139472291370951,0.0164328585136437,0.0181964406224198,0.0195643811806639,0.0206820699989684,0.0216270624299604,0.0224456521077444,0.0231676993755169,0.0238135926659885,0.0280628041513132,0.0305484335278618},yticklabels={\(\displaystyle {0}\),\(\displaystyle {10^{-3}}\),,,,,,,,,\(\displaystyle {10^{-2}}\),,},ylabel={Probability}},
colormap/viridis,
height=\figH,
point meta max=0.03068729303777,
point meta min=0,
tick align=outside,
tick pos=left,
width=\figW,
x grid style={white!69.0196078431373!black},
xlabel={Secret},
xmin=0, xmax=257,
xtick style={color=black},
y grid style={white!69.0196078431373!black},
ylabel={Time (cycles)},
ymin=91546, ymax=92580,
ytick style={color=black}
]
\addplot graphics [includegraphics cmd=\pgfimage,xmin=0, xmax=257, ymin=91546, ymax=92580] {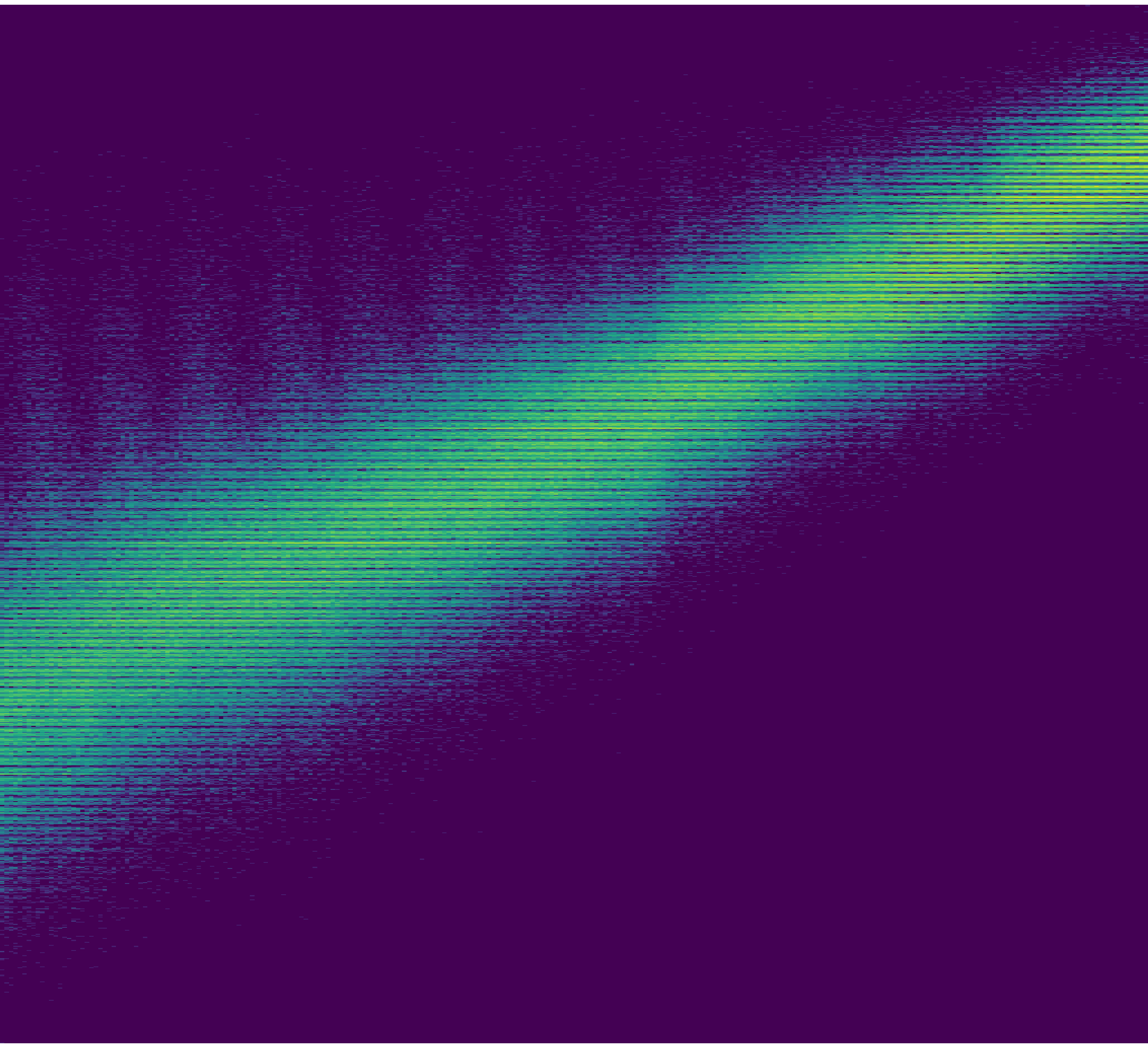};
\end{axis}

\end{tikzpicture}
        \caption{Software. (1165/0.5)}
        \label{f:cm-sw2}
    \end{subfigure}
    \\\vspace{2em}
    \begin{subfigure}[t]{\linewidth}
        \centering
        \begin{tikzpicture}[scale=\cmScale]

\definecolor{color0}{rgb}{0.267004,0.004874,0.329415}

\begin{axis}[cmStyle,
axis background/.style={fill=color0},
colorbar,
colorbar style={ytick={0,0.0437687232305772,0.0629461023741812,0.0741641500353015,0.0821234815177852,0.0882972186504182,0.0933415291789055,0.0976064329615387,0.101300860661389,0.104559576840026,0.107474597794022,0.126651976937626,0.137870024598747,0.14582935608123,0.152003093213863,0.157047403742351,0.161312307524984,0.165006735224834,0.168265451403471,0.171180472357467,0.190357851501071,0.201575899162192,0.209535230644675,0.215708967777308,0.220753278305796,0.225018182088429,0.228712609788279,0.231971325966916,0.234886346920912,0.254063726064516},yticklabels={\(\displaystyle {0}\),\(\displaystyle {10^{-4}}\),,,,,,,,,\(\displaystyle {10^{-3}}\),,,,,,,,,\(\displaystyle {10^{-2}}\),,,,,,,,,\(\displaystyle {10^{-1}}\),},ylabel={Probability}},
colormap/viridis,
height=\figH,
point meta max=0.2614795863628,
point meta min=0,
tick align=outside,
tick pos=left,
width=\figW,
x grid style={white!69.0196078431373!black},
xlabel={Secret},
xmin=0, xmax=257,
xtick style={color=black},
y grid style={white!69.0196078431373!black},
ylabel={Time (cycles)},
ymin=92521, ymax=92698,
ytick style={color=black}
]
\addplot graphics [includegraphics cmd=\pgfimage,xmin=0, xmax=257, ymin=92521, ymax=92698] {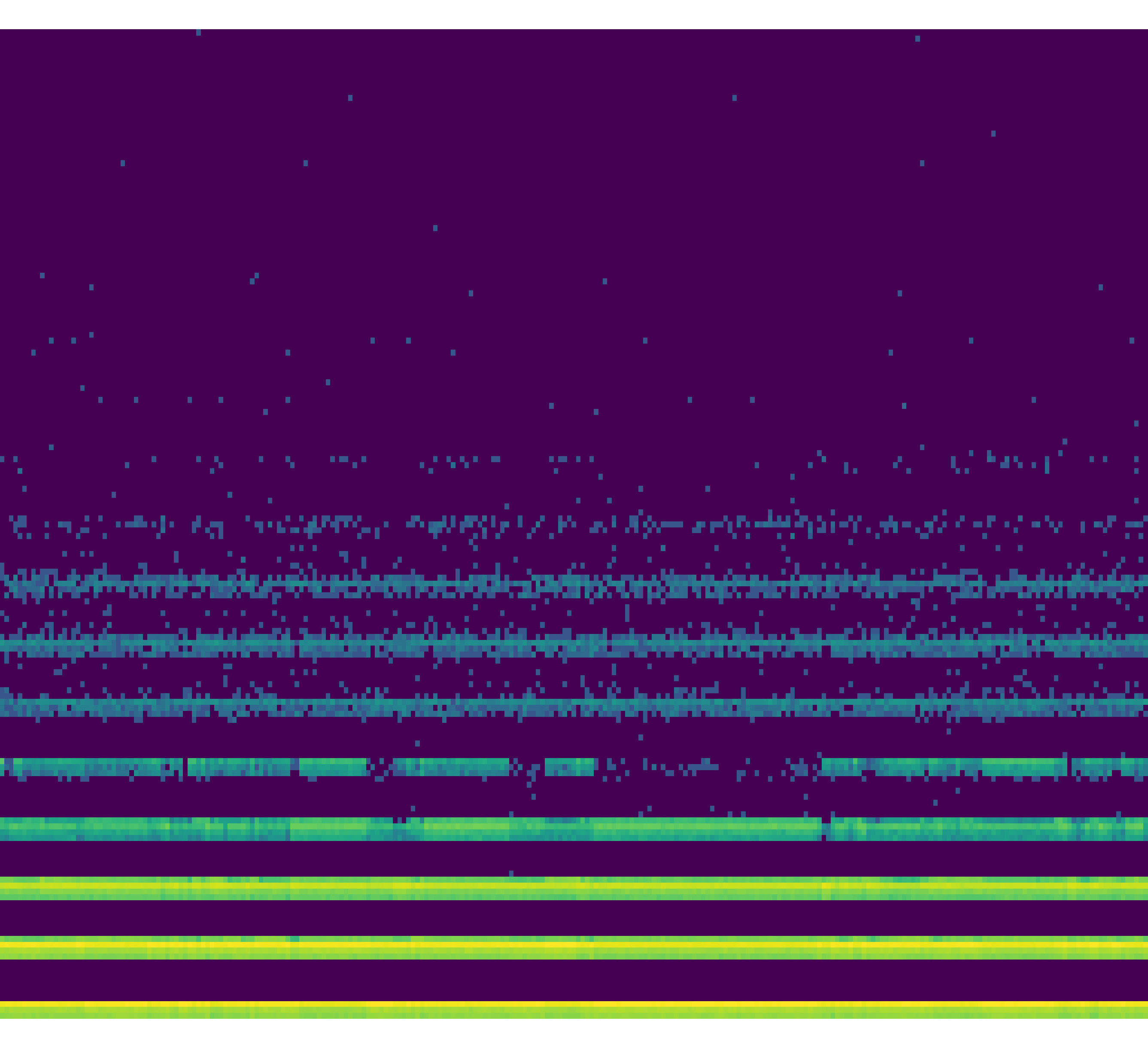};
\end{axis}

\end{tikzpicture}
        \caption{\firstflush{}. (10.7/1.6)}
        \label{f:cm-selfirst}
    \end{subfigure}
    \\\vspace{2em}
    \begin{subfigure}[t]{\linewidth}
        \centering
        \begin{tikzpicture}[scale=\cmScale]

\definecolor{color0}{rgb}{0.267004,0.004874,0.329415}

\begin{axis}[cmStyle,
axis background/.style={fill=color0},
colorbar,
colorbar style={ytick={0,0.141271293181516,0.203169675210547,0.239377907552843,0.265068057239578,0.284994885442794,0.301276289581873,0.315042020638118,0.326966439268609,0.337484521924169,0.346893267471825,0.408791649500856,0.444999881843151,0.470690031529886,0.490616859733102,0.506898263872182,0.520663994928426,0.532588413558917,0.543106496214477,0.552515241762133,0.614413623791164,0.650621856133459,0.676312005820194,0.696238834023411,0.71252023816249,0.726285969218734},yticklabels={\(\displaystyle {0}\),\(\displaystyle {10^{-3}}\),,,,,,,,,\(\displaystyle {10^{-2}}\),,,,,,,,,\(\displaystyle {10^{-1}}\),,,,,,},ylabel={Probability}},
colormap/viridis,
height=\figH,
point meta max=0.7300380468369,
point meta min=0,
tick align=outside,
tick pos=left,
width=\figW,
x grid style={white!69.0196078431373!black},
xlabel={Secret},
xmin=0, xmax=257,
xtick style={color=black},
y grid style={white!69.0196078431373!black},
ylabel={Time (cycles)},
ymin=92780, ymax=92961,
ytick style={color=black}
]
\addplot graphics [includegraphics cmd=\pgfimage,xmin=0, xmax=257, ymin=92780, ymax=92961] {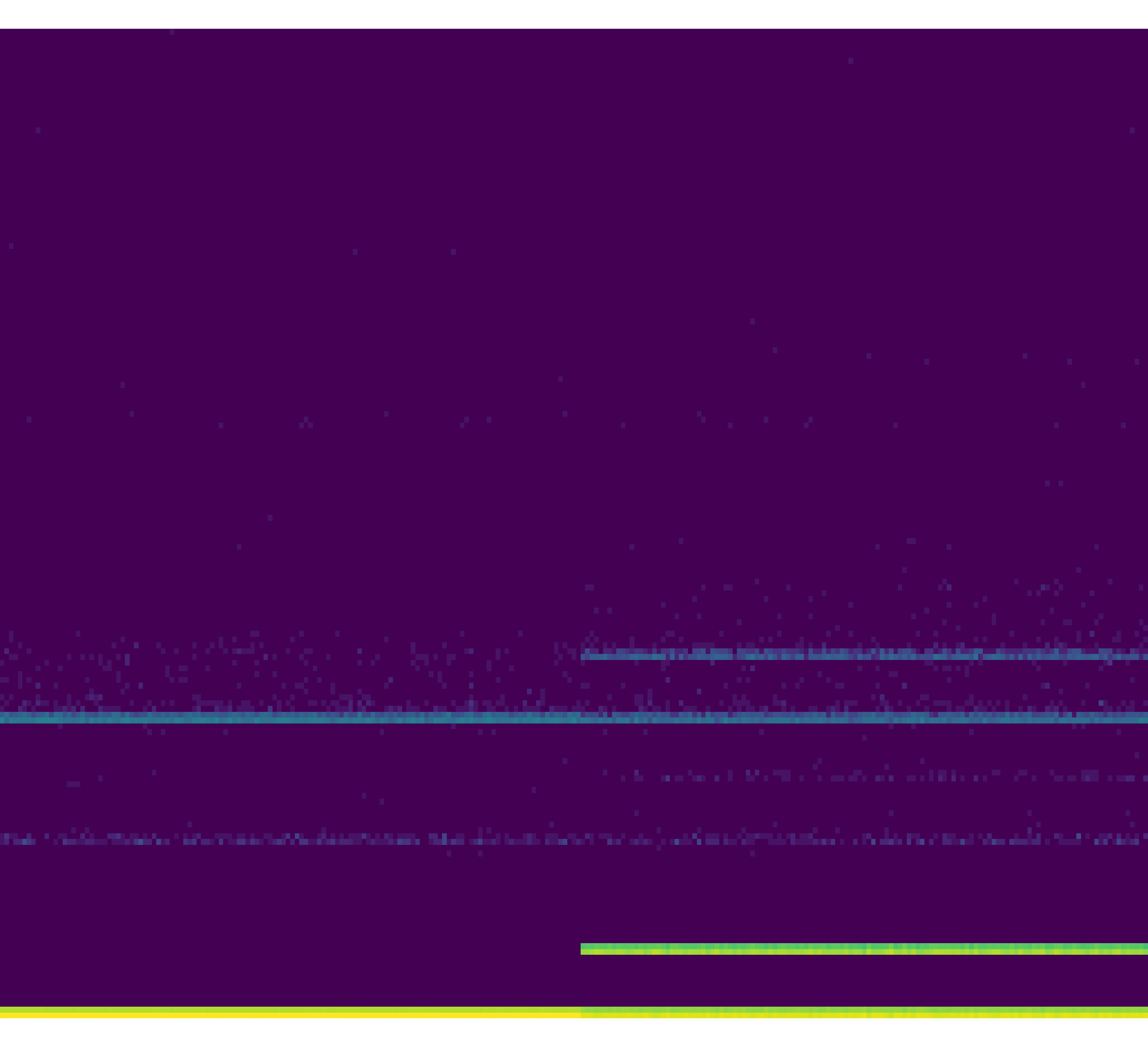};
\end{axis}

\end{tikzpicture}
        \caption{\fullflush{}. (248.4/0.1)}
        \label{f:cm-selall}
    \end{subfigure}
    \\\vspace{2em}
    \begin{subfigure}[t]{\linewidth}
        \centering
        \begin{tikzpicture}[scale=\cmScale]

\definecolor{color0}{rgb}{0.267004,0.004874,0.329415}

\begin{axis}[cmStyle,
axis background/.style={fill=color0},
colorbar,
colorbar style={ytick={0,0.105707885398016,0.152024068441943,0.179117298699178,0.198340251485871,0.213250732055656,0.225433481743105,0.235733849837306,0.244656434529798,0.252526712000339,0.259566915099584,0.305883098143511,0.332976328400745,0.352199281187438,0.367109761757224,0.379292511444672,0.389592879538873,0.398515464231365,0.406385741701907,0.413425944801151,0.459742127845078,0.486835358102312,0.506058310889005,0.520968791458791,0.53315154114624,0.543451909240441,0.552374493932932,0.560244771403474,0.567284974502718,0.613601157546645,0.64069438780388,0.659917340590572,0.674827821160358,0.687010570847807},yticklabels={\(\displaystyle {0}\),\(\displaystyle {10^{-4}}\),,,,,,,,,\(\displaystyle {10^{-3}}\),,,,,,,,,\(\displaystyle {10^{-2}}\),,,,,,,,,\(\displaystyle {10^{-1}}\),,,,,},ylabel={Probability}},
colormap/viridis,
height=\figH,
point meta max=0.6970264911652,
point meta min=0,
tick align=outside,
tick pos=left,
width=\figW,
x grid style={white!69.0196078431373!black},
xlabel={Secret},
xmin=0, xmax=257,
xtick style={color=black},
y grid style={white!69.0196078431373!black},
ylabel={Time (cycles)},
ymin=92780, ymax=92954,
ytick style={color=black}
]
\addplot graphics [includegraphics cmd=\pgfimage,xmin=0, xmax=257, ymin=92780, ymax=92954] {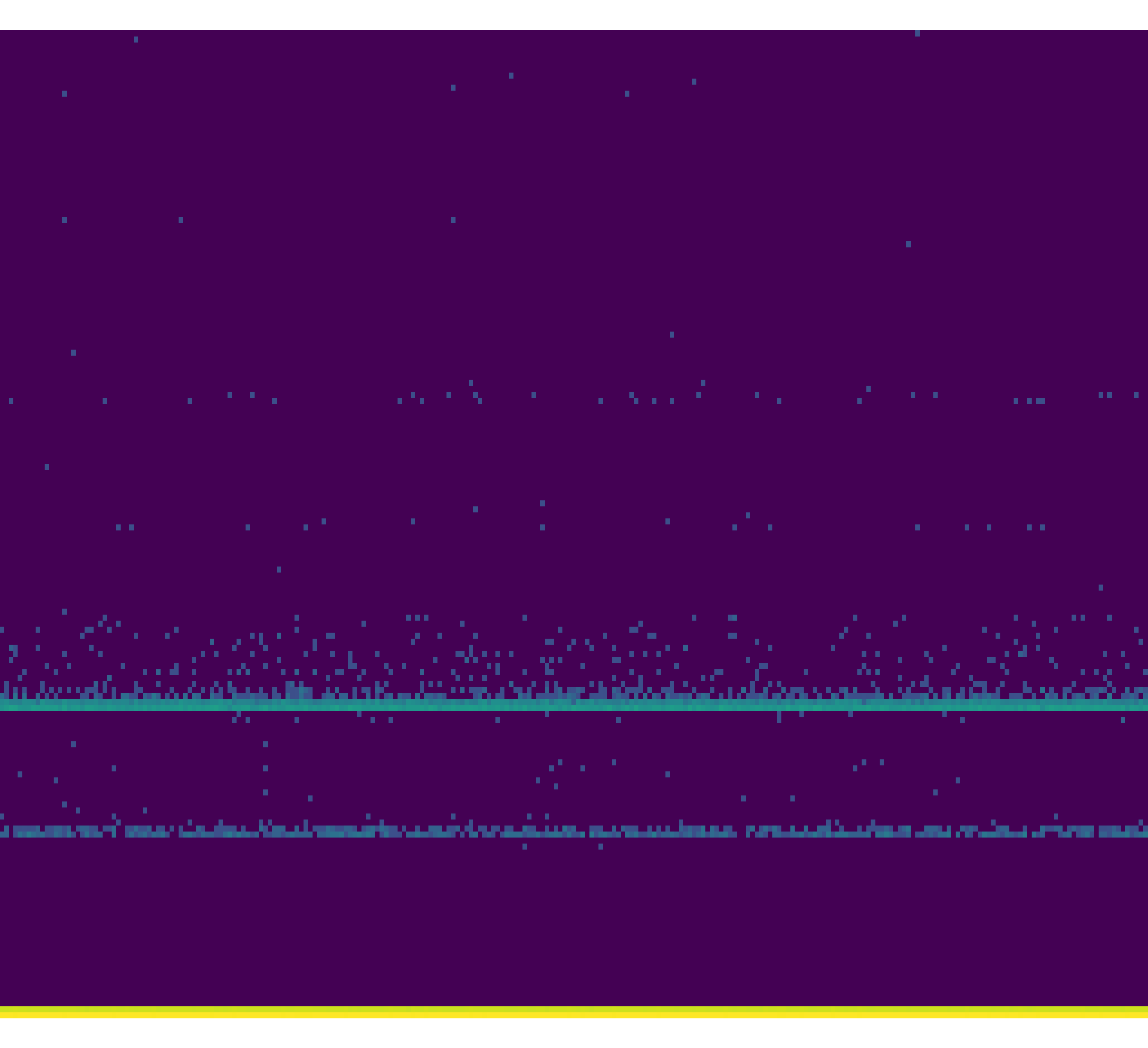};
\end{axis}

\end{tikzpicture}
        \caption{\textsc{Microreset}. (21.9/27.8)}
        \label{f:cm-urst}
    \end{subfigure}
    \caption{Channel matrices and corresponding mutual information ($\mathcal{M}$[mb]/$\mathcal{M}_0$[mb]) for the write-through L1D.}
    \label{f:cm-l1d}
\end{figure}

\begin{table*}
    \setlength\tabcolsep{0pt}
    \centering
    \begin{tabular*}{\textwidth}{%
        l                                         %
        @{\extracolsep{4ex}}S[table-format=4.0]   %
        @{\extracolsep{1ex}}S[table-format=1.1]   %
        @{\extracolsep{\fill}}c                   %
        @{\extracolsep{1ex}}c                     %
        @{\extracolsep{\fill}}S[table-format=3.1] %
        @{\extracolsep{1ex}}S[table-format=3.1]   %
        @{\extracolsep{\fill}}S[table-format=3.1] %
        @{\extracolsep{1ex}}S[table-format=3.1]   %
        @{\extracolsep{\fill}}S[table-format=3.1] %
        @{\extracolsep{1ex}}S[table-format=3.1]   %
        @{\extracolsep{4ex}}S[table-format=4.0]   %
        @{\extracolsep{1ex}}S[table-format=1.1]   %
        @{\extracolsep{\fill}}c                   %
        @{\extracolsep{1ex}}c                     %
        @{\extracolsep{\fill}}S[table-format=3.1] %
        @{\extracolsep{1ex}}S[table-format=3.1]   %
        @{\extracolsep{\fill}}S[table-format=3.1] %
        @{\extracolsep{1ex}}S[table-format=3.1]   %
        @{\extracolsep{\fill}}S[table-format=3.1] %
        @{\extracolsep{1ex}}S[table-format=3.1]}  %
        \toprule
             & \multicolumn{10}{c}{Write-Through L1} & \multicolumn{10}{c}{Write-Back L1} \\
        \cmidrule{2-11} \cmidrule{12-21}
             & \multicolumn{2}{c}{None} & \multicolumn{2}{c}{SW} & \multicolumn{2}{c}{\firstflush{}} & \multicolumn{2}{c}{\fullflush{}} & \multicolumn{2}{c}{\textsc{Microreset}} & \multicolumn{2}{c}{None} & \multicolumn{2}{c}{SW} & \multicolumn{2}{c}{\firstflush{}} & \multicolumn{2}{c}{\fullflush{}} & \multicolumn{2}{c}{\textsc{Microreset}} \\
        \cmidrule{2-3} \cmidrule{4-5} \cmidrule{6-7} \cmidrule{8-9} \cmidrule{10-11} \cmidrule{12-13} \cmidrule{14-15} \cmidrule{16-17} \cmidrule{18-19} \cmidrule{20-21}
             & \multicolumn{1}{c}{$\mathcal{M}$} & \multicolumn{1}{c}{$\mathcal{M}_0$} & \multicolumn{1}{c}{$\mathcal{M}$} & \multicolumn{1}{c}{$\mathcal{M}_0$} & \multicolumn{1}{c}{$\mathcal{M}$} & \multicolumn{1}{c}{$\mathcal{M}_0$} & \multicolumn{1}{c}{$\mathcal{M}$} & \multicolumn{1}{c}{$\mathcal{M}_0$} & \multicolumn{1}{c}{$\mathcal{M}$} & \multicolumn{1}{c}{$\mathcal{M}_0$} & \multicolumn{1}{c}{$\mathcal{M}$} & \multicolumn{1}{c}{$\mathcal{M}_0$} & \multicolumn{1}{c}{$\mathcal{M}$} & \multicolumn{1}{c}{$\mathcal{M}_0$} & \multicolumn{1}{c}{$\mathcal{M}$} & \multicolumn{1}{c}{$\mathcal{M}_0$} & \multicolumn{1}{c}{$\mathcal{M}$} & \multicolumn{1}{c}{$\mathcal{M}_0$} & \multicolumn{1}{c}{$\mathcal{M}$} & \multicolumn{1}{c}{$\mathcal{M}_0$}\\
        \midrule
        L1D  & \bfseries 1629 & 0.5 & \bfseries 1165 & 0.5 & \bfseries 10.7 & 1.6   & \bfseries 248.4 & 10.0  & 21.9  & 27.8  & \bfseries 1620 & 0.5 & \bfseries 770 & 1   & 36.0          & 36.3  & 34.4 & 36.8  & 32.6 & 37.7  \\
        L1I  & \bfseries 1891 & 0.5 & n/a            & n/a & \bfseries 9.5  & 1.4   & 33.4            & 42.0  & 42.0  & 42.5  & \bfseries 1893 & 0.5 & n/a           & n/a & \bfseries 9.1 & 3.0   & 43.0 & 48.7  & 13.5 & 13.4  \\
        DTLB & \bfseries 378  & 0.1 & n/a            & n/a & 1.7            & 4.4   & 4.8             & 5.7   & 4.3   & 7.9   & \bfseries 363  & 0.1 & n/a           & n/a & 69.0          & 90.0  & 37.6 & 91.4  & 60.9 & 91.6  \\
        BTB  & \bfseries 3611 & 0.1 & n/a            & n/a & 54.8           & 134.4 & 84.2            & 156.5 & 137.6 & 161.7 & \bfseries 3690 & 0.1 & n/a           & n/a & 85.2          & 158.2 & 92.3 & 181.3 & 83.1 & 162.7 \\
        BHT  & \bfseries 3933 & 0.4 & n/a            & n/a & 137.4          & 160.5 & \bfseries 161.0 & 160.0 & 0.0   & 0.0   & \bfseries 4147 & 0.2 & n/a           & n/a & 118.8         & 167.2 & 99.8 & 162.2 & 0.0  & 0.0   \\
        \bottomrule
    \end{tabular*}
    \caption{Timing channel capacities and their corresponding zero-leakage upper bounds for the unmitigated design and the discussed mitigation mechanisms in millibit [mb]. Leaking channels are highlighted.}
    \label{t:channels}
\end{table*}

\autoref{f:cm-l1d} shows the result of \textsc{Channel Bench} for the write-through L1 data cache for different implementation approaches of time protection.
We use the write-through L1 data cache as an example for an in-depth security analysis and comparison of the proposed mechanisms.
Most of the following observations also hold for the other microarchitectural components.

\paragraph{Unmitigated}
As a baseline, we use the original, unmodified CVA6 core and run our testbench without any further on-core time protection in seL4.
\autoref{f:cm-base} shows the resulting channel matrix.
A clear correlation between the Trojan's secret and the spy's execution time is visible, indicating the presence of a covert channel.
This is confirmed by the mutual information $\mathcal{M}$, which is clearly above the zero-leakage upper bound $\mathcal{M}_0$ at more than \SI{1.6}{\bit} per iteration.
To illustrate this channel's bandwidth, let us assume a 256-bit AES key, two concurrently running applications, and a time slice of \SI{1}{\milli\second}.
The AES key could be leaked in less than \SI{320}{\milli\second}.
More efficient encodings could achieve even higher throughput.

\paragraph{Mitigation Using Existing Architecture}
We next evaluate the approach of \ref{s:sw}, using only existing instructions to mitigate the timing channel.
As the results in \autoref{f:cm-sw2} and \autoref{t:channels} show, this decreases the channel's capacity without fully closing it.
One explanation for this behaviour lies within the replacement policy of the data cache.
CVA6 pseudo-randomly selects a cache entry for eviction in case of a collision.
As a result, the \gls{os} cannot reliably evict all data cache entries on a context switch.
It is possible to re-iterate the prime sequence, but the security guarantees remain limited, and the performance costs increase rapidly, as shown in \autoref{s:costs-cs}.
We conclude that the current architecture does not provide the \gls{os} sufficient means to enforce time protection, and hardware support is needed.

\paragraph{Basic Flush (\firstflush{})}
\label{s:eval-selfirst}
The channel matrix for the basic flush, presented in \autoref{s:selfirst}, is shown in \autoref{f:cm-selfirst}.
While the overall appearance of the channel matrix is flat, some patterns along the x-axis remain.
Additionally, the mutual information is clearly above the zero-leakage upper bound, confirming a residual channel.
A closer analysis reveals that the timing of memory accesses is not only determined by the state of the cache itself, but also by that of further stateful components, such as the \gls{lfsr} providing a pseudo-random index sequence for the cache replacement policy, and the round-robin memory arbiters of the core.
Concurrently to our work, Vila et al.~\cite{Vila2020Flushgeist} made similar observations on an Intel core.

\paragraph{Full Flush (\fullflush{})}
\label{s:full-flush}
The full flush (\autoref{s:selall}) clears these \emph{secondary} components as well.
While we close most channels with this approach, sporadically, a binary channel such as the one shown in \autoref{f:cm-selall} reappears in the write-through L1 data cache.
The channel does not exist consistently for each measurement.
We observe that it appears depending on the initial hardware state.

Running \textsc{Channel Bench} on CVA6 in RTL simulation, we find that a single-cycle flush of the targeted components is insufficient.
As CVA6 is a pipelined design, the flush signal may reach the various components at different points in time.
If the components are not reset synchronously, information can flow from one component that is not yet reset to another component that has already been reset and thus persist.
A possible approach to solving this issue is to apply the flush signal for multiple cycles to ensure it propagates through the whole design before being de-asserted, as we do for \textsc{Microreset}. We do not explore this path further for the full flush.

Another channel that was identified through this analysis is the miss handler of the L1 data cache.
When it receives a new request just before \fencet{} is executed, it waits for the cache's write-back and flush procedure to complete before serving the request.
The response is discarded later on, but it still leaves a trace on state that was already reset by \fencet{}.
This trace depends on the request that was issued before \fencet{}, and therefore previous execution.

Although these channels might appear very small and impractical at first sight, they become more prominent as additional sources of noise are removed with the reset of other components.
We see this as the main reason for sporadically higher channel capacities of \fullflush{} compared to \firstflush{} shown in \autoref{t:channels}.
Also, it is important to highlight that our L1 data cache attack shown in \autoref{f:cm-l1d} does not target these channels specifically---they are only visible as a side effect of the L1 data cache attack.
An attack directly targeting the presented channels would, presumably, achieve much higher capacity.

\paragraph{Microreset}
Finally, \textsc{Microreset} from \autoref{s:urst} yields the expected result, as demonstrated in \autoref{f:cm-urst}.
The channel is consistently closed across configuration and attacks, as supported by \autoref{t:channels}.

\subsubsection{Further Components}

Besides the L1 data cache, we analyse prime-and-probe attacks on the L1 instruction cache, the data \gls{tlb}, the \gls{btb}, and the \gls{bht}, see \autoref{t:channels}.
They confirm our findings from the L1 data cache: the unmodified design leaks significant amounts of data (e.g. more than \SI{4}{\bit} per iteration for the \gls{bht}), while executing \fencet{} using \textsc{Microreset} during a context switch reliably closes all channels.

\subsubsection{Context-Switch Latency}
\label{s:eval-cs}

\begin{figure}
    \centering
    \def\BarHeight{1cm}

\begin{tikzpicture}[x=0.01\linewidth, y=0.01\linewidth]
    \node (spy1)   [minimum height=\BarHeight, minimum width=1cm,   align=center]                               {\small Spy};
    \node (cs1)    [minimum height=\BarHeight, minimum width=1.8cm, align=center, anchor=west] at (spy1.east)   {\small Context\\Switch};
    \node (trojan) [minimum height=\BarHeight, minimum width=3cm,   align=center, anchor=west] at (cs1.east)    {\small Trojan};
    \node (cs2)    [minimum height=\BarHeight, minimum width=1.8cm, align=center, anchor=west] at (trojan.east) {\small Context\\Switch};
    \node (spy2)   [minimum height=\BarHeight, minimum width=1cm,   align=center, anchor=west] at (cs2.east)    {\small Spy};
    \draw [-] (spy1.north west)   -- (spy2.north east);
    \draw [-] (spy1.south west)   -- (spy2.south east);
    \draw [-] (spy1.north east)   -- (spy1.south east);
    \draw [-] (cs1.north east)    -- (cs1.south east);
    \draw [-] (trojan.north east) -- (trojan.south east);
    \draw [-] (cs2.north east)    -- (cs2.south east);

    \node (clint1) [anchor=south, align=center] at ($(cs1.north west)+(0,3)$) {\small \textit{CLINT}\\\textit{interrupt}};
    \node (clint2) [anchor=south, align=center] at ($(cs2.north west)+(0,3)$) {\small \textit{CLINT}\\\textit{interrupt}};
    \draw [-] (clint1.south) -- (cs1.north west);
    \draw [-] (clint2.south) -- (cs2.north west);

    \draw [dotted] (spy1.south east) -- ++(0,-5);
    \draw [dotted] (spy2.south west) -- ++(0,-5);
    \draw [-stealth] ($(spy1.south east)-(0,5)$) -- ($(spy2.south west)-(0,5)$);
\end{tikzpicture}
    \caption{Time span measured by the spy in the context-switch latency channel.}
    \label{f:cs-overview}
\end{figure}
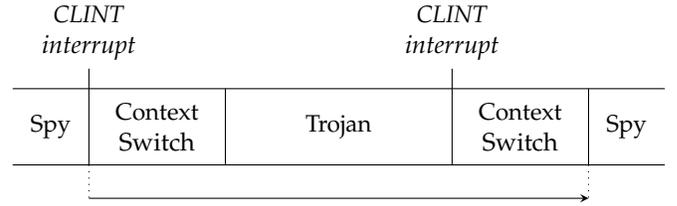

\setlength{\figW}{0.6\linewidth}
\setlength{\figH}{\figW}

\begin{figure*}
    \centering
    \begin{subfigure}[t]{0.33\linewidth}
        \centering
        \begin{tikzpicture}[scale=\cmScale]

\definecolor{color0}{rgb}{0.267004,0.004874,0.329415}

\begin{axis}[cmStyle,
axis background/.style={fill=color0},
colorbar,
colorbar style={ytick={0,0.00236027175021071,0.00472054350042142,0.00708081525063213,0.00836845975077223,0.00936723441942883,0.0101832927433345,0.0108732602250784,0.0114709372434746,0.0119981257358968,0.0124697119121312},yticklabels={\(\displaystyle {0}\),\(\displaystyle {10^{-3}}\),,,,,,,,,\(\displaystyle {10^{-2}}\)},ylabel={Probability}},
colormap/viridis,
height=\figH,
point meta max=0.01396431401372,
point meta min=0,
tick align=outside,
tick pos=left,
width=\figW,
x grid style={white!69.0196078431373!black},
xlabel={Secret},
xmin=0, xmax=257,
xtick style={color=black},
y grid style={white!69.0196078431373!black},
ylabel={Time (cycles)},
ymin=402702, ymax=403736,
ytick style={color=black}
]
\addplot graphics [includegraphics cmd=\pgfimage,xmin=0, xmax=257, ymin=402702, ymax=403736] {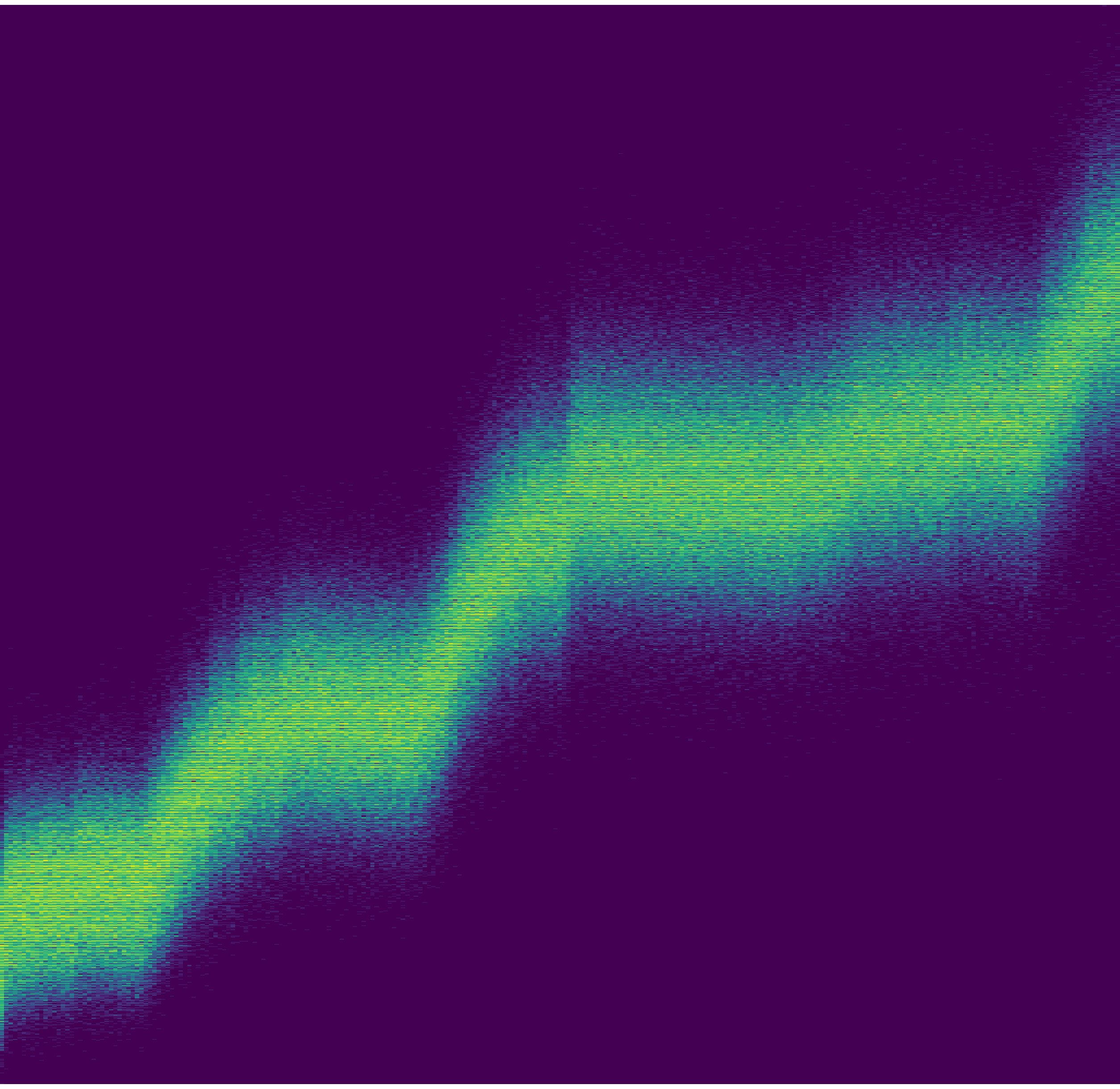};
\end{axis}

\end{tikzpicture}
        \small
        \begin{align*}
            M &= \SI{1297.9}{\milli\bit} \\
            M_0 &= \SI{0.5}{\milli\bit}
        \end{align*}
        \caption{Unmitigated.}
        \label{f:cm-cs}
    \end{subfigure}
    \hfill
    \begin{subfigure}[t]{0.33\linewidth}
        \begin{tikzpicture}[scale=\cmScale]

\definecolor{color0}{rgb}{0.267004,0.004874,0.329415}

\begin{axis}[cmStyle,
axis background/.style={fill=color0},
colorbar,
colorbar style={ytick={0,0.0779852012537635,0.1121546186287,0.132142446464528,0.146324036003637,0.157324131442562,0.166311863839465,0.173910883305133,0.180493453378574,0.186299691675293,0.191493548817498,0.225662966192435,0.245650794028263,0.259832383567372,0.270832479006296,0.2798202114032,0.287419230868868,0.294001800942309,0.299808039239028,0.305001896381233,0.33917131375617,0.359159141591998},yticklabels={\(\displaystyle {0}\),\(\displaystyle {10^{-3}}\),,,,,,,,,\(\displaystyle {10^{-2}}\),,,,,,,,,\(\displaystyle {10^{-1}}\),,},ylabel={Probability}},
colormap/viridis,
height=\figH,
point meta max=0.3694202303886,
point meta min=0,
tick align=outside,
tick pos=left,
width=\figW,
x grid style={white!69.0196078431373!black},
xlabel={Secret},
xmin=0, xmax=257,
xtick style={color=black},
y grid style={white!69.0196078431373!black},
ylabel={Time (cycles)},
ymin=405661, ymax=423063,
ytick style={color=black}
]
\addplot graphics [includegraphics cmd=\pgfimage,xmin=0, xmax=257, ymin=405661, ymax=423063] {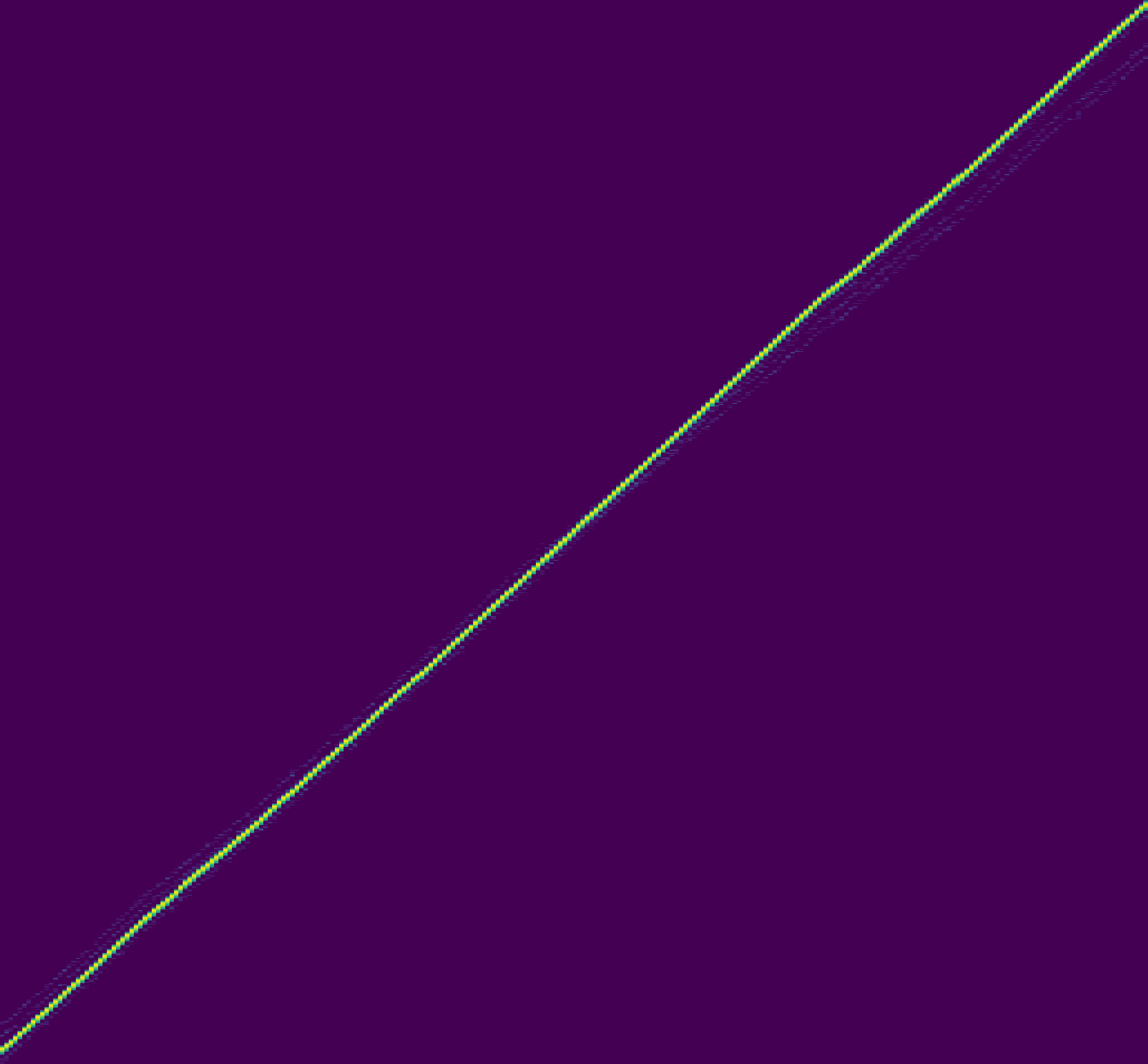};
\end{axis}

\end{tikzpicture}
        \small
        \begin{align*}
            M &= \SI{7257.2}{\milli\bit} \\
            M_0 &= \SI{0.4}{\milli\bit}
        \end{align*}
        \caption{\fencet{} \emph{without} time padding.}
        \label{f:cm-cs-fencet}
    \end{subfigure}
    \hfill
    \begin{subfigure}[t]{0.33\linewidth}
        \begin{tikzpicture}[scale=\cmScale]

\definecolor{color0}{rgb}{0.267004,0.004874,0.329415}

\begin{axis}[cmStyle,
axis background/.style={fill=color0},
colorbar,
colorbar style={ytick={0,0.0213883077656956,0.0307596500619289,0.0362415338866474,0.0401309923581621,0.0431478907301258,0.0456128761828807,0.0476969916873126,0.0495023346543954,0.0510947600075992,0.0525192330263591,0.0618905753225923,0.0673724591473109,0.0712619176188256,0.0742788159907893,0.0767438014435441,0.0788279169479761,0.0806332599150588},yticklabels={\(\displaystyle {0}\),\(\displaystyle {10^{-3}}\),,,,,,,,,\(\displaystyle {10^{-2}}\),,,,,,,},ylabel={Probability}},
colormap/viridis,
height=\figH,
point meta max=0.08076131343842,
point meta min=0,
tick align=outside,
tick pos=left,
width=\figW,
x grid style={white!69.0196078431373!black},
xlabel={Secret},
xmin=0, xmax=257,
xtick style={color=black},
y grid style={white!69.0196078431373!black},
ylabel={Time (cycles)},
ymin=424253, ymax=424655,
ytick style={color=black}
]
\addplot graphics [includegraphics cmd=\pgfimage,xmin=0, xmax=257, ymin=424253, ymax=424655] {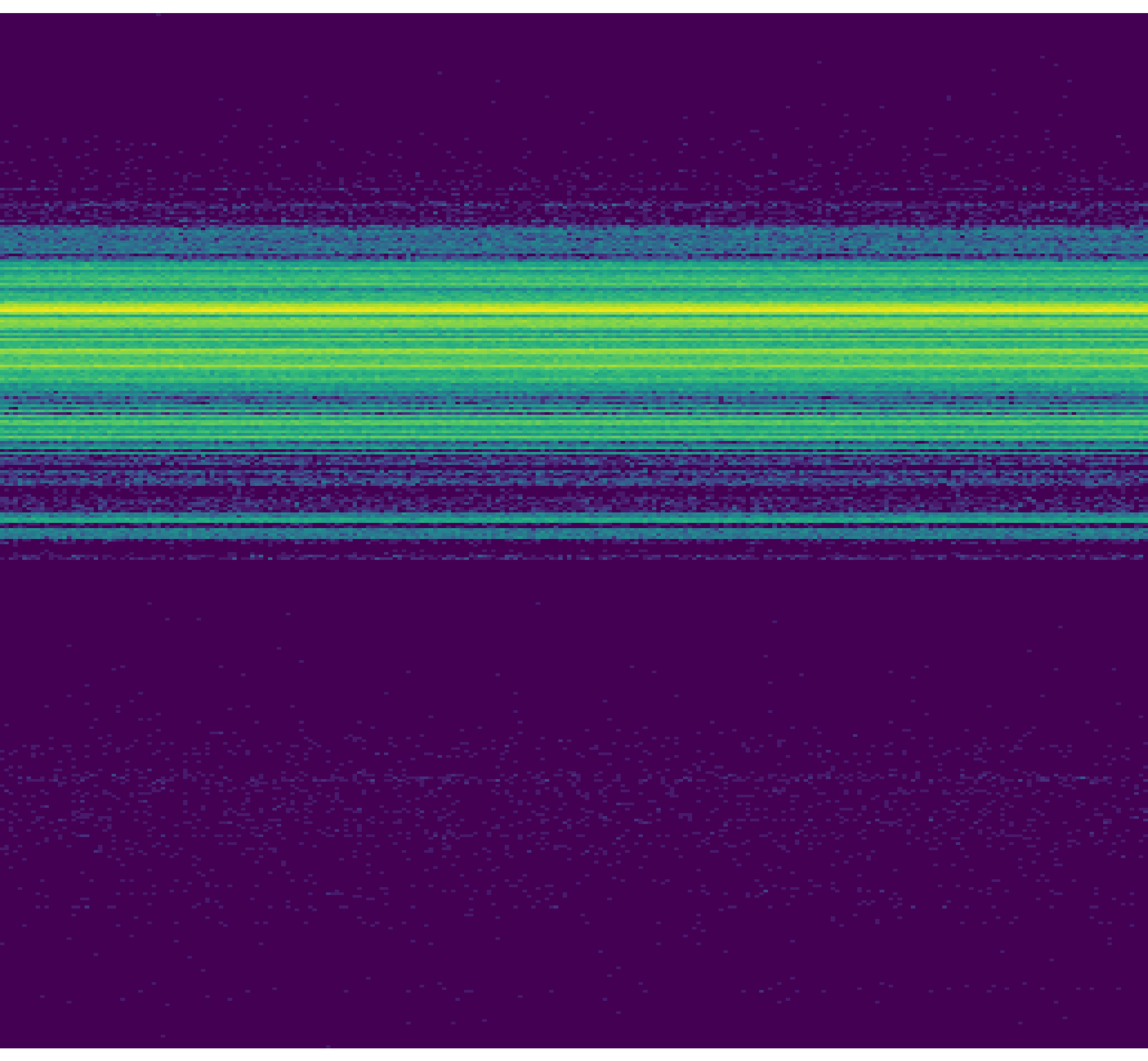};
\end{axis}

\end{tikzpicture}
        \small
        \begin{align*}
            M &= \SI{1.4}{\milli\bit} \\
            M_0 &= \SI{1.6}{\milli\bit}
        \end{align*}
        \caption{\fencet{} \emph{with} time padding.}
        \label{f:cm-cs-fencet-pad}
    \end{subfigure}
    \caption{Evicted time measured by spy over secret number of L1 data cache lines written by Trojan.}
\end{figure*}

To evaluate leakage through the context-switch latency, we configure the spy to measure the time span during which it is evicted, as shown in \autoref{f:cs-overview}.
In this interval, the Trojan application runs for a fixed time-slice before the \gls{os} performs a context switch back to the spy.

\autoref{f:cm-cs} shows the secret number of L1 data cache lines that the Trojan writes on the horizontal axis and the corresponding eviction duration measured by the spy on the vertical axis for the unmitigated case.
A clear correlation between both is visible, indicating a covert channel.
When the Trojan writes to cache lines, it evicts any kernel data stored at the same location.
Thus, the context switch routine causes cache misses, increasing the context switch latency.

Resetting the microarchitectural state by executing \fencet{} on a context switch \emph{without} accounting for the latency worsens the situation: \autoref{f:cm-cs-fencet} shows a covert channel close to its theoretical upper capacity limit of \SI{8}{\bit}.
As \fencet{} needs to write back the L1 data cache's dirty cache lines sequentially, its latency directly depends on the secret number of previously written cache lines.

We proposed padding for the worst-case execution latency in \autoref{s:pad}.
Hence, we measure the latency for a fully dirty write-back cache: it takes $3\,077\,(\pm 7)$ cycles from the \gls{clint} timer interrupt until the start of the execution of \fencet{} execution.%
\footnote{The majority of this latency (approximately 1800 cycles) are spent for reconfiguring the \gls{clint}.
Scheduling takes around 800 cycles and switching to the new thread (e.g.~changing the address space) takes another 320 cycles.}
It then takes $18\,646\,(\pm 4)$ cycles for writing back and invalidating the dirty L1 data cache.
Finally, $16\,(\pm 0)$ cycles are spent draining pending transactions, and \textsc{Microreset} is asserted for another $16\,(\pm 0)$ cycles, resulting in a total of $21\,755\,(\pm 8)$ cycles from the \gls{clint} timer interrupt until the end of \fencet{}.
We conservatively round up and set $\cspad{} = 22\,000$.
Similarly, we determine a worst-case upper bound of 3\,700 cycles for the write-through L1 data cache.
As shown in \autoref{f:cm-cs-fencet-pad}, this removes any dependence of the context switch latency on previous execution and microarchitectural state.

\subsection{Costs}

\subsubsection{Context-Switch Latency}
\label{s:costs-cs}

For evaluating the context switch latency, we use the inter-address-space IPC benchmark from sel4bench~\cite{github:sel4bench}.
Analogue to \autoref{s:pad}, we set $\cspad{} = 22\,000$.
However, as the benchmark uses the process-initiated \emph{fastpath} context switch routine, in contrast to \autoref{s:pad}, the beginning of the context switch routine is not defined by a \gls{clint} timer interrupt.
Hence, for our performance evaluation, we use the privilege level switch from U-mode as the start of the pad interval.
This event approximates the \gls{clint} timer interrupt for full context switches.

\begin{table}
    \setlength\tabcolsep{0pt}
    \centering
    \begin{tabular*}{\columnwidth}{
        l
        @{\extracolsep{2ex}}l
        @{\extracolsep{\fill}}m{1.5em}
        @{\extracolsep{0ex}}S[table-format=5.0]
        @{ $\pm$ }S[table-format=1.0]
        m{0em}
        @{\extracolsep{\fill}}m{0.5em}
        @{\extracolsep{0ex}}S[table-format=5.0]
        @{ $\pm$ }S[table-format=3.0]
        m{0em}}
        \toprule
        \multicolumn{2}{l}{\multirow{2}{*}{Mitigation}}  & \multicolumn{4}{c}{Write-Through L1} & \multicolumn{4}{c}{Write-Back L1} \\
        \cmidrule{3-6} \cmidrule{7-10}
                                      &                  && {Mean}           & {SD}             &&& {Mean}        & {SD}           &\\
        \midrule
        \multirow{2}{*}{None}      & Hot                 && 514              & 0                &&& 423           & 0              &\\
                                   & Cold/Dirty          && 1243             & 2                &&& 1827          & 217            &\\[1ex]
        SW                         &                     && 40650            & 2                &&& 41152         & 15             &\\[1ex]
        \multirow{3}{*}{\fencet{}} & \firstflush{}       && 4073             & 1                &&& 22402         & 5              &\\
                                   & \fullflush{}        && 4067             & 1                &&& 22402         & 5              &\\
                                   & \textsc{Microreset} && 4125             & 0                &&& 22450         & 0              &\\
        \bottomrule
    \end{tabular*}
    \caption{
        seL4 fastpath context switch latency in cycles without mitigation, mitigated using existing architecture (SW), and with \fencet{}.
        \fencet{} is padded for a worst-case slowpath context switch.
    }
    \label{t:cs}
\end{table}

\autoref{t:cs} shows the context-switch latencies for different configurations.
\emph{Hot} assumes that the core very recently executed a context switch; hence the caches contain kernel data, the branch predictors are trained, etc.
No on-core mitigations against timing channels are in place.
On the other hand, \emph{Cold/Dirty} is the context-switch latency for an untrained and dirty microarchitecture.
For instance, this is the case after a resource-intensive application has evicted all kernel information during its time slice.
We assume that this is the more common case in practice.
We emulate this behaviour by resetting the microarchitecture and, if configured as write-back, polluting the L1 data cache from user space between context switches.
Finally, \emph{SW} and \fencet{} are the resulting context-switch latencies after configuring seL4 to mitigate on-core timing channels using the approaches described in \autoref{s:impl}.

The Software mitigation (\emph{SW}) is significantly more expensive than the \fencet{} approaches while only partially mitigating the L1 data cache channel.
A more extensive mitigation would make this approach even more costly.

Comparing \textsc{Microreset} to both flush approaches, there are no significant performance differences.
The reason is that most components of CVA6 are reset in parallel---therefore, flushing or resetting more state usually does not require more cycles.
The dominating factor for both approaches is the write-back of the L1 data cache padded for the worst case.
This means that a principled reset generally does not imply higher costs than a selective flush.

Compared to the cold, unmitigated case, \fencet{} adds less than 21\,000 cycles to the context-switch routine.
Assuming a processor running at \SI{1}{\giga\hertz} and a context-switch frequency of \SI{100}{\hertz}, this increase corresponds to an overhead of about \SI{0.2}{\percent}.
Decreasing the frequency of \fencet{} (e.g.\ in a hypervisor scenario or by clustering mutually trusting applications into security domains) would decrease the relative overhead accordingly.
It is important to note that \fencet{} is padded for the worst-case slowpath context switch, while \emph{Cold/Dirty} gives the fastpath latency.
When applied to a slowpath context switch, the overhead would be smaller.
In addition to the direct costs shown here, the dirty cache would cause indirect costs resulting from cache misses and write-backs experienced after the context switch, while after \fencet{}, execution continues with a clean cache.
As such, \autoref{t:cs} overestimates the performance impact of \fencet{}.

\subsubsection{Indirect Costs}

Resetting the microarchitectural state on a context switch potentially removes an application's information from the on-core state that is still required at a later point, resulting in indirect costs due to cache misses or mis-speculation once the application is re-scheduled.
However, Ge et al. have shown that application-specific information is mostly evicted from the on-core state (L1 caches, branch predictors etc.) after one or several time-slices of execution of other application(s), and that the indirect costs of a reset on context switch are therefore limited.

A difference in indirect costs between \textsc{Microreset} and the selective flush can generally only be caused either by a poor choice of reset values for one of both approaches, or by a residual timing channel.

\subsubsection{Benchmark results}
\label{s:bm}

We evaluate the overhead introduced by \fencet{} with \textsc{Microreset} using the Splash-2 benchmarks~\cite{Woo1995splash}.
We set up two domains: the first executing the benchmark and the second concurrently running an idle thread.
We perform a context switch between both domains every 10 million cycles, which corresponds to a typical \SI{10}{\milli\second} timeslice period on a processor running at \SI{1}{\giga\hertz}.

\autoref{f:bm} shows the slowdown of the benchmark when executing a \fencet{} on every context switch.
The standard error is below \SI{0.03}{\percent} for all results.
The average slowdown is \SI{0.7}{\percent}, which confirms the low performance overhead of \fencet{}.

We note that this is a rather pessimistic evaluation: the idle thread has no significant memory footprint, meaning that for the baseline, the benchmarks benefit from a hot microarchitecture.
In a setting with high memory contention, the latency of \fencet{} (dominated by the cache flush) may be almost entirely hidden.

\begin{figure}
    \centering
    \small
    \begin{tikzpicture}
    \begin{axis}[
        width  = \linewidth,
        height = 4cm,
        major x tick style = transparent,
        ybar=2*\pgflinewidth,
        bar width=6pt,
        ymajorgrids = true,
        ylabel = {Slowdown},
        symbolic x coords={barnes,cholesky,fft,ffm,lu,ocean,radiosity,radix,raytrace,waternsquared,waterspatial},
        xtick = data,
        xtick align=outside,
        xtick style={},
        xticklabel style={rotate=45,anchor=east,xshift=3pt,yshift=-3pt},
        scaled y ticks = false,
        yticklabel={\pgfmathprintnumber\tick\%},
        enlarge x limits=0.05,
        ymin=0,
        legend cell align=left,
        legend style={
                at={(0.98,0.95)},
                anchor=north east,
                column sep=1ex
        }
    ]
        \addplot[fill=black,style={mark=none}]
            coordinates {(barnes, 0.30) (cholesky, 0.54) (fft, 0.54) (ffm, 0.50) (lu, 1.58) (ocean, 0.43) (radiosity, 0.26) (radix, 0.21) (raytrace, 0.46) (waternsquared, 0.09) (waterspatial, 0.47)};

        \addplot[pattern=north east lines,style={mark=none}]
            coordinates {(barnes, 0.12) (cholesky, 3.52) (fft, 0.86) (ffm, 0.45) (lu, 1.11) (ocean, 0.99) (radiosity, 0.02) (radix, 1.07) (raytrace, 0.40) (waternsquared, 0.18) (waterspatial, 0.49)};

        \legend{write-back L1,write-through L1}
    \end{axis}
\end{tikzpicture}
    \caption{Splash-2 benchmark slowdown by \fencet{}}
    \label{f:bm}
\end{figure}

\subsubsection{Hardware Overhead}

We synthesise the original and modified versions of CVA6 in \textsc{GlobalFoundries} 22 FDX technology at \SI{1}{\giga\hertz} at worst-case conditions (\SI{0.72}{\volt}, \SI{125}{\celsius}).
We convert the results to gate equivalent (GE), a technology-independent unit for the complexity of a circuit.
The area overhead of our modifications is negligible at 0.4\%, with the \fencet{} controller being the largest addition at around \SI{1.6}{\kilo\GE} compared to a total core area of \SI{1.2}{\mega\GE}.
There is no significant impact on the critical path.

\section{Related Work}
\label{s:related-work}

The L1 data cache is the focus of several previous works on on-core timing channel mitigation.
One possible approach is spatial partitioning of the cache, proposed by Page~\cite{Page2005Partition} and followed up on by Domnitser et al.~\cite{Domnitser2012NoMo} and Dessouky et al.~\cite{Dessouky2021ChunkedCache}.
Since L1 caches are relatively small and time-shared between applications, spatial partitioning of these components is not very efficient.
Wang and Lee~\cite{Wang2007DynamicCache1, Wang2008DynamicCache2} propose a dynamic, randomised cache-remapping to mitigate targeted cache collisions.
An improved implementation was presented by Qureshi~\cite{Qureshi2018CEASER}.
We find this approach insufficient, as in general, randomisation merely adds noise to a communication channel without fundamentally closing it.
As Constable and Unterluggauer~\cite{Constable2021SEED} demonstrate, cache line mappings be designed to prevent specific attacks (i.e. prime-and-probe on the index of an accessed cache line).
Besides imposing impractical constraints on the cache layout, this approach is not suited for other attacks, such as prime-and-probe on the \emph{number} of accessed cache lines, which we use in this work.

All works mentioned above do not consider microarchitectural components besides the L1 data cache.
Tiwari et al.~\cite{Tiwari2009ExecutionLeases} propose a major architectural modification to fundamentally separate data from control flow.
Our work extends that of Ge et al., who propose time protection and the need for flushing all microarchitectural on-core state on a partition switch, and demonstrate the need for hardware support~\cite{Ge2018,Ge2019a,Ge2019phd}, which is what our temporal fence provides.
There exist several approaches in a similar direction:
Bourgeat et al. \cite{Bourgeat2019MI6} present a processor with a \texttt{purge} instruction, similar to our \fencet{}, that flushes on-core microarchitectural components to secure enclaves.
Li et al.~\cite{Li2020SIMF} propose \textsc{FenceX}, an instruction similar to the basic flush version of \fencet{} presented in \autoref{s:selfirst} of this work, which we found insufficient to close all timing channels reliably.
Escouteloup et al.~\cite{Escouteloup2021Dome} present an \gls{isa} extension that allows the allocation of hardware resources to security domains.
They implement and evaluate their proposal bare-metal on an embedded RISC-V core designed to model known microarchitectural vulnerabilities, whereas this work targets an existing, application-class RISC-V core, optimised for efficiency and running a full operating system.

To the best of our knowledge, all previous works on temporal partitioning first identify vulnerable microarchitectural components, and then add them to a partition set.
A major drawback of this approach is that it remains difficult to make hard claims, such as that \emph{all} microarchitectural covert channels are closed.
Our \textsc{Microreset} proposed in \autoref{s:urst} works the other way around: all stateful components are reset per default.
Only a selected set of architectural state is explicitly excluded from the reset.
Furthermore, we also consider the flush latency itself, since neglecting it can open significant new channels, as demonstrated in \autoref{s:eval-cs}.

\section{Conclusions}
\label{s:conclusions}
In this work, we present the temporal fence instruction, \fencet{}, which allows an \gls{os} to reliably prevent on-core timing channels.
We propose and compare different hardware implementations of \fencet{}, ranging from an basic flush of well-known vulnerable microarchitectural components, over an exhaustive flush of manually identified vulnerable secondary components, to a systematic erasure of all non-architectural state, which we call \textsc{Microreset}.
Evaluating these mechanisms on the open-source RV64GC CVA6 core with different cache configurations and running an experimental, unverified version of the seL4 microkernel, we find that \fencet{} with \textsc{Microreset} is the only approach that consistently closes all timing channels, while offering a low implementation effort, a performance impact of less than \SI{1}{\percent} for a typical \SI{1}{\giga\hertz} system with a \SI{10}{\milli\second} context switch period, and negligible hardware costs.

\ifCLASSOPTIONcompsoc
  \section*{Acknowledgments}
\else
  \section*{Acknowledgment}
\fi

The authors would like to thank Qian Ge and Curtis Millar for their support with the covert channel measurement framework, Florian Zaruba for his help and insights on CVA6, and Paul Scheffler for his feedback on the manuscript.

The work of Wistoff, G\"urkaynak, and Benini was supported in part by the European Union's Horizon 2020 research and innovation programme FRACTAL project funded by ECSEL-JU grant agreement \#877056, ETH4HC, and the ETH4D Humanitarian Action Challenges Application on “Secure Infrastructure for Humanitarian Organizations".

\ifCLASSOPTIONcaptionsoff
  \newpage
\fi

\bibliographystyle{IEEEtran}
\bibliography{IEEEabrv,sources}

\newpage

\begin{IEEEbiography}[{\includegraphics[width=1in,height=1.25in,clip,keepaspectratio]{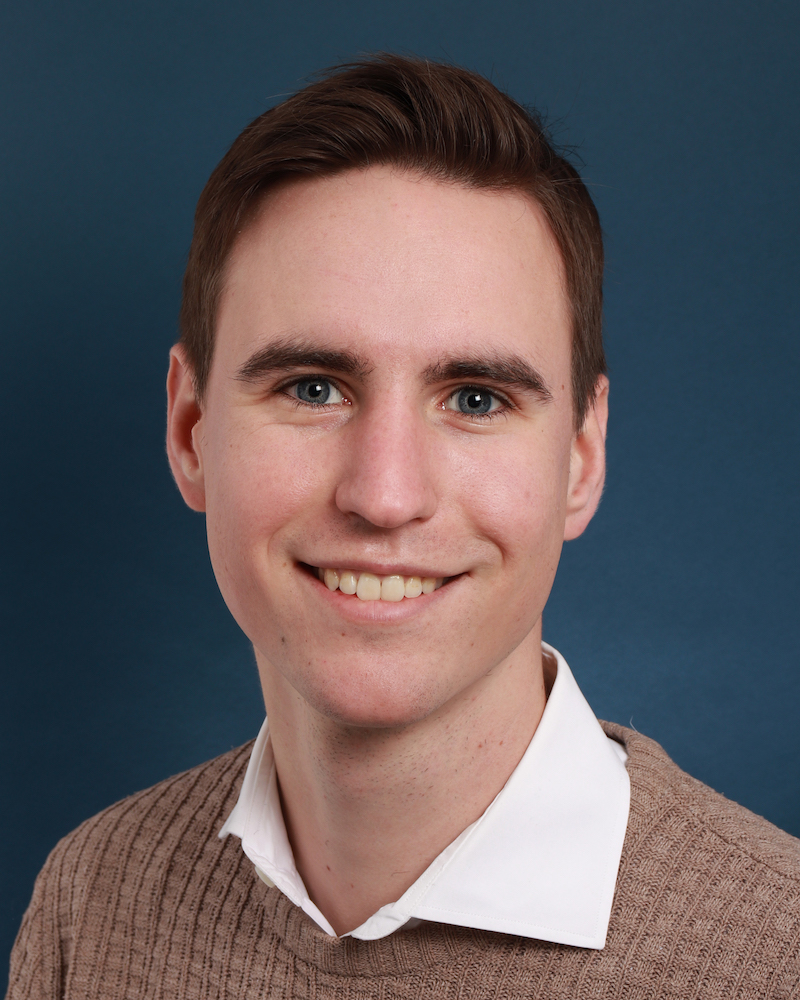}}]{Nils Wistoff}
received his B.Sc. and M.Sc. from RWTH Aachen University in 2017 and 2020, respectively.
He is currently pursuing a Ph.D. at the Integrated Systems Laboratory of ETH Z\"urich.
Wistoff's research interests include processor and system-on-chip design and secure computer architecture.
\end{IEEEbiography}

\vfill

\begin{IEEEbiography}[{\includegraphics[width=1in,height=1.25in,clip,keepaspectratio]{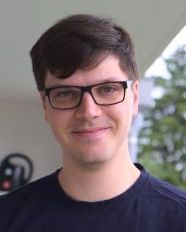}}]{Moritz Schneider}
has received his B.Sc. and M.Sc. in electrical engineering from ETH Zurich and he is currently pursuing a Ph.D. in the System Security group at ETH Zurich. His research interests include hardware security, trusted execution, and system security.
\end{IEEEbiography}

\vfill

\begin{IEEEbiography}[{\includegraphics[width=1in,height=1.25in,clip,keepaspectratio]{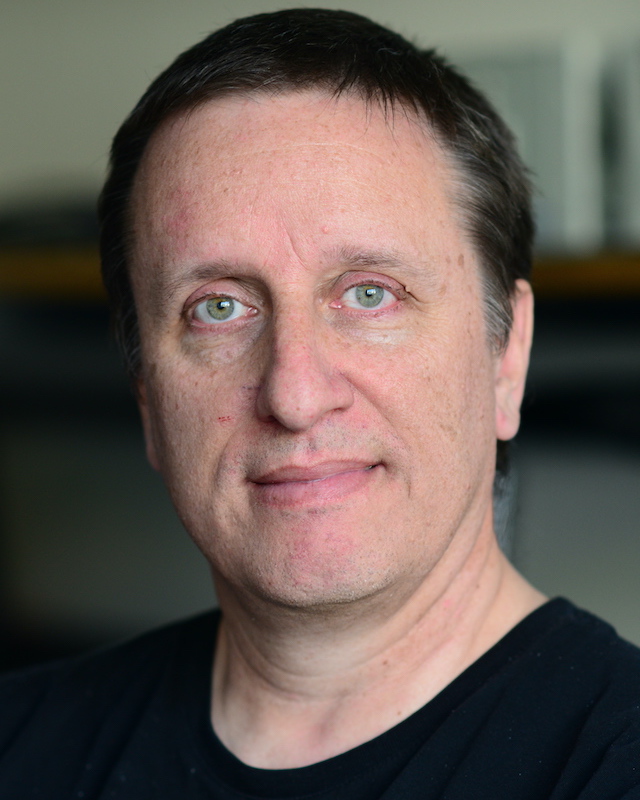}}]{Frank K. G\"urkaynak}
has obtained his B.Sc. and M.Sc. in electrical engineering from the Istanbul Technical University, and his Ph.D. in electrical engineering from ETH Z\"urich in 2006. He is currently working as a senior scientist at the Integrated Systems Laboratory of ETH Z\"urich. His research interests include digital low-power design and cryptographic hardware.
\end{IEEEbiography}

\vfill

\begin{IEEEbiography}[{\includegraphics[width=1in,height=1.25in,clip,keepaspectratio]{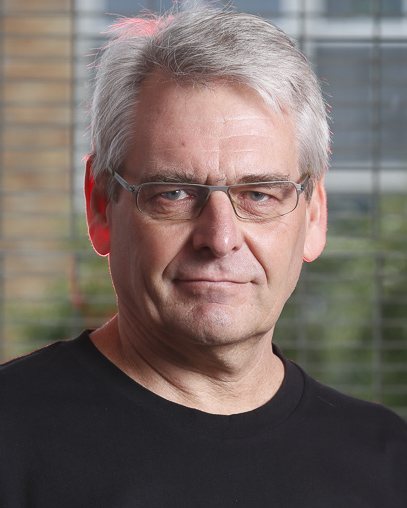}}]{Gernot Heiser}
is Scientia Professor and John Lions Chair at UNSW Sydney. His
research interests are provably secure and dependable operating
systems, and their application in security- and safety-critical uses
cases. He also serves as the chairman of the seL4 Foundation. Dr
Heiser is a Fellow of the ACM and the Australian Academy of Technology
and Engineering (ATSE) and a winner of the ACM SIGOPS Hall of Fame award.
\end{IEEEbiography}

\vfill

\begin{IEEEbiography}[{\includegraphics[width=1in,height=1.25in,clip,keepaspectratio]{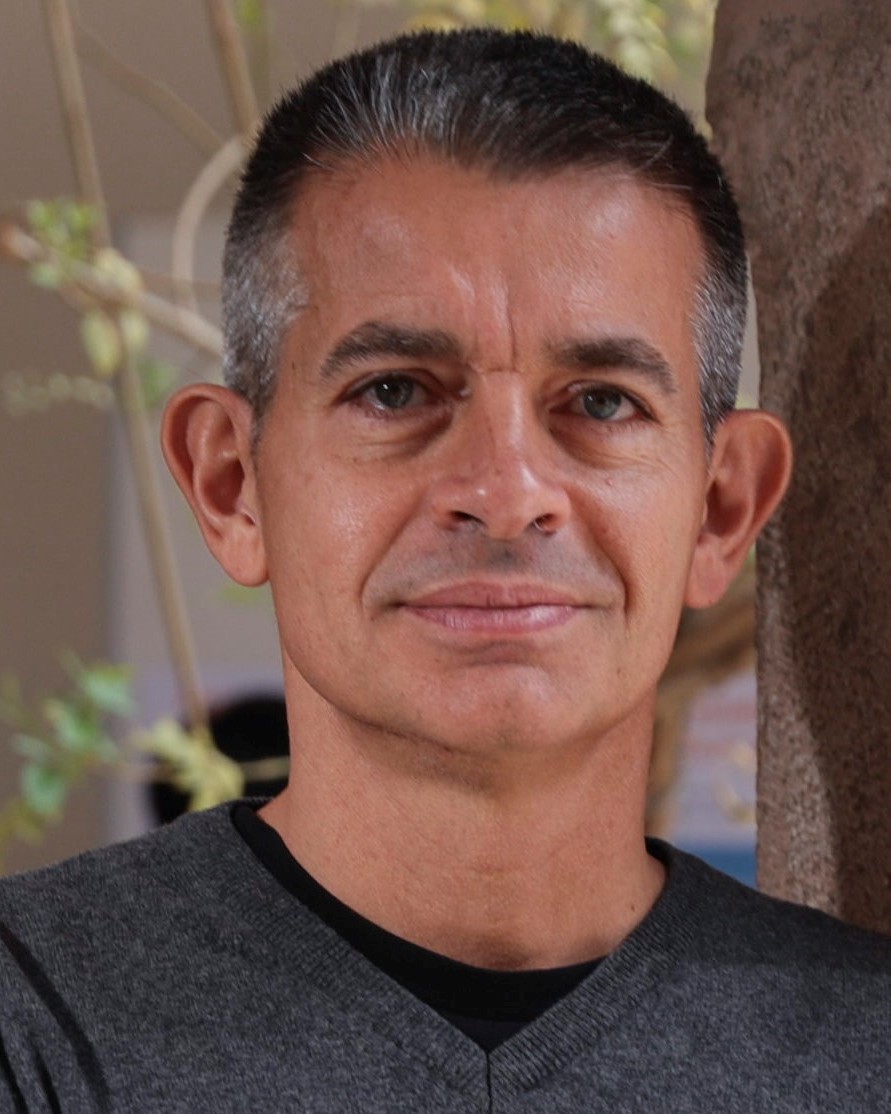}}]{Luca Benini}
holds the chair of digital Circuits and systems at ETHZ and is Full Professor at the Universita di Bologna.
Dr. Benini’s research interests are in energy-efficient computing systems design, from embedded to high-performance.
He has published more than 1000 peer-reviewed papers and five books.
He is a Fellow of the ACM and a member of the Academia Europaea.
He is the recipient of the 2016 IEEE CAS Mac Van Valkenburg award.
\end{IEEEbiography}

\end{document}